\documentclass[fleqn,usenatbib]{mnras}

\usepackage{newtxtext,newtxmath}
 
\usepackage[T1]{fontenc}
\usepackage{ae,aecompl}

\usepackage{graphicx}	
\usepackage{amsmath}	
\usepackage{amssymb}	
\usepackage{xcolor}
\usepackage[labelfont=bf]{caption}

\newcommand{\kmsmpc}{\kms\;{\rm Mpc}^{-1}}

\newcommand{\hkpc}{h^{-1}{\rm kpc}}
\newcommand{\hmpc}{h^{-1}{\rm Mpc}}

\newcommand{\kms}{\;{\rm km}\,{\rm s}^{-1}}

\newcommand{\gizmo}{{\sc Gizmo}}
\newcommand{\mufasa}{{\sc Mufasa}}
\newcommand{\simba}{{\sc Simba}}

\newcommand{\fgas}{f_{\rm gas}}
\newcommand{\fedd}{f_{\rm Edd}}
\newcommand{\mbh}{\;{\rm M}_{\rm BH}}
\newcommand{\mstar}{\;{\rm M}_{*}}
\newcommand{\mgas}{\:{\rm M}_{\rm gas}}
\newcommand{\ro}{R_{0}}
\newcommand{\msun}{\;{\rm M}_{\odot}}
\newcommand{\mhalo}{\;{\rm M}_{\rm halo}}
\newcommand{\mdot}{\;{\dot{\rm M}}_{\rm BH}}

\newcommand{\WHz}{\rm{~W~Hz}^{-1}}
\newcommand{\power}{\;{\rm P}_{\rm 1.4GHz}}
\newcommand{\edit}{\textcolor{black}}
\newenvironment{Figure}
  {\par\medskip\noindent\minipage{\linewidth}}
  {\endminipage\par\medskip}

\title[The Radio Galaxy Population in {\sc SIMBA}]{The Radio Galaxy Population in the {\sc SIMBA} Simulations}

\author[Thomas et al.]{
Nicole Thomas,$^{1,2}$\thanks{E-mail: thomas.nicolelynn@gmail.com}
Romeel Dav\'e,$^{1,2,3}$
Matt J.~Jarvis,$^{1,4}$
Daniel Angl\'es-Alc\'azar$^{5,6}$
\\
$^{1}$Department of Physics and Astronomy, University of the Western Cape, Bellville, 7535, South Africa\\
$^{2}$South African Astronomical Observatories, Observatory, Cape Town 7925, South Africa\\
$^{3}$Institute for Astronomy, Royal Observatory, University of Edinburgh, Edinburgh, EH9 3HK, UK\\
$^{4}$University of Oxford, Denys Wilkinson Building, Keble Road, Oxford, OX1 3RH, UK\\
$^{5}$Department of Physics, University of Connecticut, 196 Auditorium Road, U-3046, Storrs, CT 06269-3046, USA \\
$^{6}$Center for Computational Astrophysics, Flatiron Institute, 162 Fifth Avenue, New York, NY 10010, USA
}

\date{Accepted XXX. Received YYY; in original form ZZZ}

\pubyear{2020}

\begin{document}
\label{firstpage}
\pagerange{\pageref{firstpage}--\pageref{lastpage}}
\maketitle

\begin{abstract}
We examine the 1.4GHz radio luminosities of galaxies arising from star formation and active galactic nuclei~(AGN) within the state-of-the-art cosmological hydrodynamic simulation \simba. \simba\ grows black holes via gravitational torque limited accretion from cold gas and Bondi accretion from hot gas, and employs AGN feedback including jets at low Eddington ratios.
We define a population of radio loud AGN (RLAGN) based on the presence of ongoing jet feedback. Within RLAGN we define high and low excitation radio galaxies (HERGs and LERGs) based on their dominant mode of black hole accretion: torque limited accretion representing feeding from a cold disk, or Bondi representing advection-dominated accretion from a hot medium. 
\simba\ predicts good agreement with the observed radio luminosity function~(RLF) and its evolution, overall as well as separately for HERGs and LERGs.
Quiescent galaxies with AGN-dominated radio flux dominate the RLF at $\ga 10^{22-23}$~W~Hz$^{-1}$, while star formation dominates at lower radio powers.
Overall, RLAGN have higher black hole accretion rates and lower star formation rates than non-RLAGN at a given stellar mass or velocity dispersion, but have similar black hole masses.
\simba\ predicts a LERG number density of 8.53~Mpc$^{-3}$, $\sim 10\times$ higher than for HERGs, broadly as observed.
While LERGs dominate among most massive galaxies with the largest black holes and HERGs dominate at high specific star formation rates, they otherwise largely populate similar-sized dark matter halos and have similar host galaxy properties.  \simba\ thus predicts that deeper radio surveys will reveal an increasing overlap between the host galaxy demographics of HERGs and LERGs.

\end{abstract}




\section{Introduction}
Essentially all massive galaxies host a supermassive black hole (SMBH) at their centers~\citep{Magorrian1998,KH2013} reaching millions to billions of solar masses ($\msun$).  When they accrete matter they become ``active'', and are identified as active galactic nuclei (AGN).  The impact of SMBHs on their host galaxies and environs is a key question in galaxy evolution today, and AGN represent our clearest observational view into this process.

AGN demonstrate a wide range of properties.  Studies have forwarded several ways of classifying AGN depending on observations across the electromagnetic spectrum.  The popular AGN unification model of \citet{UrryPadovani1995} suggests that many disparate types of AGN can be unified by appealing to orientation of the line of sight relative to the obscuring circum-black hole torus, but this remains controversial~\citep{Netzer2015}.  An orthogonal AGN classification scheme proposed by \citet{BestandHeckman2012} appeals to their accretion efficiency, or Eddington fraction ($\fedd$), defined as the black hole accretion rate (BHAR) divided by the Eddington rate at its black hole mass ($\mbh$).  Effectively, $\fedd$ is a measure of the specific growth rate of SMBHs. Observationally, black holes that accrete at $\fedd\lesssim 1\%$  are typically identified as ``jet mode'' AGN due to the observed presence of relativistic jets that can span thousands of times the size of the host galaxy itself, while those that accrete at $\fedd\gtrsim 1\%$ are identified as ``radiative mode'' AGN owing to the immense amounts of radiative energy produced in the accretion disk around the SMBH~\citep{HeckmanAndBest2014}.

The host galaxies of AGN in these modes show clear differences.  Radiative mode AGN (also referred to as ``quasar mode'', ``cold mode'', or ``high excitation'' AGN) are typically associated with ongoing star formation activity within the host galaxy~\citep{Kauffmann2003} and are thought to accrete cold gas efficiently via an optically thick, geometrically thin accretion disk \citep{Shakura1973} often surrounded by an obscuring dusty torus~\citep{RamosAlmeida2017}. These AGN emit their energy efficiently across a broad range of the electromagnetic spectrum, and a fraction ($\sim 10\%$) of them can be seen emitting strongly in the radio by means of strong radio jets. These so-called powerful radio galaxies (if torus obscured and radio-loud quasars if unobscured) can be identified by high excitation lines in their optical spectra, and are thus referred to as as high excitation radio galaxies~\citep[HERGs;][]{Buttiglione2010}.

The second mode of AGN activity, jet mode (also ``radiatively inefficient'', ``hot mode'', or ``low excitation'') AGN, has black holes that accrete at an inefficient rate in such a way that leads to little radiated energy across the electromagnetic spectrum. Instead, such radiatively inefficient AGN emit most of their energy as kinetic energy in the form of powerful relativistic radio jets~\citep{Merloni2007,Whittam2018}. Jet mode AGN show little to no presence of excitation lines within their optical spectra and are thus classified as low excitation radio galaxies (LERGS). These are believed to accrete their matter from a surrounding hot gas  reservoir within the innermost regions of the galaxy comparatively slowly via advection-dominated accretion~\citep{Hardcastle2007}.

It is widely reasoned that in order for the SMBH environments of HERGs and LERGs to differ, they likely live in host galaxies with different properties. Observationally, LERGs tend to reside in the most massive galaxies that have little to no ongoing star formation, whereas HERGs reside in moderate to massive galaxies \citep{Best2007} that still have some on going star formation and a sufficient cold gas supply that can be accreted by the AGN. This ties in with the idea that HERGs require more cold gas surrounding their SMBH, albeit on scales much smaller than the entire host galaxy.

Early studies found that LERGs dominated the radio luminosity function (RLF) at low luminosities while HERGs dominated at high luminosities, but more recent studies suggest that HERGs and LERGs are found across all radio luminosities~\citep{Whittam2018}.  The radio power of LERGs are often dominated by the lobes of the jets being emitted, \edit{ while \citet{Whittam2016} found that at 15.7GHz,  HERGs have significant radio emission coming from the core}, although powerful jets and lobes can still occur (e.g. Cygnus A) \edit{and tend to dominate at lower frequencies}.  This is an indication that the LERGs typically host stronger jets relative to their accretion rate, while the radio emission in HERGs can remain within the host galaxy, potentially contaminated by star formation processes. 

Furthermore, the evolution between HERGs and LERGs are distinct, even over the relatively modest redshift range that they have been explored. \citet{BestandHeckman2012} show that HERGs evolve strongly at all radio luminosities, while LERGs show little to no evolution up to $z\approx0.3$.  Thus the evolution of the total radio luminosity function (RLF) of AGN is driven by the evolution of these two populations. \citet{Best2014} further showed that similarly, radiatively efficient radio loud AGN (RLAGN) evolve by a constant factor of $\approx7$ up to $z\approx0.75$, and the radiatively inefficient jet mode AGN show little to no evolution for luminosities $\power<10^{25}\WHz$ in agreement with \cite{Pracy2016} and \cite{Prescott2016}; however, at higher luminosities, they evolve similarly to that of the radiatively efficient AGN.  These again suggest that HERGs and LERGs occupy different host galaxy types.

The HERG/LERG classification has taken on particular significance of late in the galaxy evolution community, owing to the growing consensus that radio jets are a key actor in quenching massive galaxies~\citep{Bower2006,Croton2006,Somerville2008}. This paradigm asserts that radio mode (LERG) AGN provide energy input into the circum-galactic gas of massive galaxies, thereby counteracting cooling and eventually starving the galaxy of fuel for star formation~\citep{McNamara2007}.  Meanwhile, quasar mode (HERG) AGN may be connected to a phase associated with morphological transformation via merging~\citep{Springel2005,Hopkins2008}.  While heuristic models following this paradigm are broadly successful at reproducing the observed galaxy population when implemented into semi-analytic or hydrodynamic galaxy formation simulations~\citep{Somerville2015}, the exact connection between these types of radio galaxies and the various phases of galaxy evolution remains unclear.

\edit{The use of semi-analytic models (SAMs) have provided significant insights into modelling the growth and feedback of black holes}~\citep{Fanidakis2011, Griffin2019, Raouf2019} and have been successful in reproducing observed black hole -- galaxy scaling relations. Using the SAM {\sc SAGE},~\citet{Raouf2017} models the formation and evolution of radio jets and estimates the 1.4GHz radio luminosity of the jet cocoon by assuming the equipartition between particle and magnetic field density. The authors consider both ``hot'' and ``cold'' mode accretion based on a separation of $\mdot<\alpha_{\rm crit} \dot{\rm M}_{\rm Edd}$ and  $\mdot>\alpha_{\rm crit} \dot{\rm M}_{\rm Edd}$ respectively with $\alpha_{\rm crit}=0.03$. Advection-dominated accretion flows and thin disc jets are generated at this transition.
This model matches the stellar mass -- radio luminosity relation as $z\sim0$ and reproduces the evolution of the radio luminosity function out to $z\sim1$.

Modern cosmological-scale hydrodynamic simulations all implement a sub-grid model for black hole growth and associated feedback, though these models can differ substantively.  Virtually all simulations model SMBH growth via Bondi accretion~\citep{BondiAndHoyle,Bondi1952,Hoyle1939} which describes the spherical advection of hot gas surrounding a black hole. This would be more in line with the accretion mode of LERGs, or jet-mode AGN.  However, such models typically employ the total gas density in the Bondi accretion rate, not just the hot gas.  This enables them to also model the quasar mode during mergers when the central cold gas density peaks~\citep{Springel2005b}, but in this situation, it is less clear that the Bondi model is physically appropriate.  For instance, angular momentum loss is generally regarded as the limiting factor in feeding black holes from cold gas in a disk.  \citet{HopkinsQuataert2011} studied this problem using analytic modeling and ultra-high resolution isolated galaxy simulations, and showed that the Bondi formula is a poor representation for the accretion rate in this scenario, confirmed by recent sub-pc resolution simulations of QSO fueling in a full cosmological context \citep{DAA2020}. Instead of Bondi accretion, \citet{HopkinsQuataert2011} proposed a different sub-grid prescription that better accounts for feeding via disk instabilities.
\citet{DAA2013,DAA2015,DAA2017} implemented their model, which they dubbed ``torque-limited accretion", into cosmological-scale simulations, and showed that it predicted qualitatively different behaviour than Bondi accretion, including not requiring self-regulation by the black hole in order to grow its mass in accord with the $\mbh-\sigma$ relation.  The self-regulation requirement has been key in driving most feedback models, which are generally driven to employ a spherical or quasi-spherical black hole feedback model in order to self-regulate the BH growth~\citep{Vogelsberger2014,Schaye2015,Springel2018}.  

The Horizon-AGN simulation~\citep{Dubois2012,Dubois2014,Volonteri2016} stands apart from this mainstream in that it uses a more complicated accretion model along with bipolar feedback, more in accordance with observed jet feedback.  Although it did not yield a massive quenched galaxy population in accord with observations, they nonetheless examined the radio galaxy population in this model.  \citet{Slyz2015} estimated the RLF within Horizon-AGN for both star formation and AGN contributions up to $z=4$, and find consistency with~\citet{MS2007} observations. For the AGN contribution to the RLF, the authors select varying Eddington fractions $f_{\rm Edd,low}$ and $f_{\rm Edd,high}$ such that for galaxies with $\fedd<f_{\rm Edd,low}$, their radio power is estimated using equation~\ref{eq:rad_all} in~\S\ref{sec:RLAGN} describing low luminosity AGN, while for $\fedd>f_{\rm Edd,high}$ the radio power is estimated via equation~\ref{eq:rad_hergs} describing radio loud AGN, interpolated in between.  With appropriate choices of $f_{\rm Edd,low}$ and $f_{\rm Edd,high}$, they could broadly reproduce the observed RLF at a range of redshifts.  This demonstrated that cosmological simulations could be used to study the nature of radio galaxies, but the somewhat arbitrary parameter choices and lack of a direct connection between jets and accretion modes argue for a more physically based model for radio AGN.

The \simba\ cosmological hydrodynamic simulation~\citep{Dave2019} differs substantively from others in its subgrid black hole accretion model.  \simba\ is the only model to introduce a dichotomy in the accretion modes, employing
torque-limited accretion for cold gas and Bondi accretion for hot gas.  This naturally maps onto the HERG and LERG accretion modes envisioned by \citet{HeckmanAndBest2014}.  Moreover, because torque-limited accretion does not require black holes to self-regulate their own growth, it is able to employ purely bipolar feedback, following the observational guidance that high-$\fedd$ black holes eject highly mass-loaded winds at moderate velocities while low-$\fedd$ black holes eject jets at high velocities.  This scheme has been quite successful at reproducing a wide range of observations of galaxy~\citep{Dave2019,Appleby2019,Dave2020,Glowacki2020}, black hole~\citep{Thomas2019,Habouzit2020}, and intergalactic gas~\citep{Christiansen2020,Sorini2020}. \simba\ thus uniquely has all the ingredients to not only study radio galaxies but also to physically classify them as HERGs or LERGs, albeit necessarily via subgrid models.  This potentially opens up new avenues for understanding the role of radio galaxies within the overall galaxy population, and constraining galaxy formation models based on radio continuum observations.


Upcoming radio surveys aim to substantially expand the samples of radio galaxies both nearby and at higher redshifts.
For example, the MeerKAT International GigaHerts Tiered Extragalactic Exploration (MIGHTEE,~\citealt{Jarvis2017}) is a recently-started radio continuum survey observing $\sim20$ square degrees of the southern sky using MeerKAT, a precursor to the Square Kilometer Array (SKA) radio interferometer. This will be complemented by ample multi-wavelength data in its well-observed fields, enabling unprecedented characterisation of radio galaxies within the overall galaxy population.
A key science goal of MIGHTEE is thus to understand AGN accretion and feedback within the ensemble of galaxies, by probing AGN and star formation activity back to Cosmic Noon using obscuration-free radio data.  Another SKA precursor survey, the LOFAR Two-Meter Sky Survey~\citep[LoTSS;][]{LOFAR2019} is an ongoing high resolution, low frequency survey of the entire northern sky providing a large sample of star-forming galaxies and RLAGN~\citep{Hardcastle2019}.  Given the emergence of these and other multi-wavelength radio surveys, it is timely to employ cosmological models to understand the nature of radio galaxies.

In this work, we use \simba\ to investigate the 1.4 GHz radio luminosity properties of simulated galaxies.  The goal is to first determine whether \simba\ yields a plausible population of radio galaxies, and then to provide some basic characterisations into how radio galaxies relate to the underlying galaxy population.
To do this, we employ literature models for computing the radio emission owing to
star formation and AGN from predicted BH and stellar properties, and compare the resulting radio luminosity functions versus observations from today back to $z\sim 3$. We differentiate between HERGs and LERGs via their dominant accretion mode as outlined above, and show that \simba\ separately yields good agreement with the observed demographics of these sub-populations.  Finally, we examine the host galaxy properties of radio galaxies as well as where these populations lie on galaxy scaling relations, as a step towards contextualising present and upcoming radio galaxy surveys.  Overall, we find good agreement with available observations, while revealing interesting predictions that can be tested with upcoming radio surveys.  This provides a first step towards employing RLAGN as a constraint for cosmological galaxy formation models, and understanding their role in establishing the present-day galaxy population.

This paper is organised as follows. In \S\ref{sec:sims} we discuss the \simba\ simulations and the black hole accretion (\S\ref{sec:accretion}) and feedback (\S\ref{sec:feedback}) models implemented therein. In \S\ref{sec:radio_em} we describe the radio emission prescriptions. In \S\ref{sec:RGs}, \ref{sec:Evolve}, and \ref{sec:host_props} we present the resulting RLFs, the large scale properties of HERGs and LERGs and their evolution, respectively. We summarise our findings in \S\ref{sec:conclusions}.

\section{Simulations}
\label{sec:sims}
\simba~\citep{Dave2019} is a state-of-the-art suite of cosmological hydrodynamic simulations which \edit{makes use of} a branched version of the \gizmo\ cosmological gravity and hydrodynamics solver~\citep{Hopkins2015} in its Meshless Finite Mass (MFM) version. It is based on much of the framework described by the \mufasa\ simulations~\citep{Dave2016} with a few notable adjustments to improve physical realism. In particular, the modelling of black hole growth and feedback is a new addition to \simba.  Because of its relevance to the present work, we will detail the modelling of black holes within \simba, and only briefly recap the other aspects; a full description is available in \citet{Dave2019}.

We use \simba's fiducial $(100 \hmpc)^{3}$ box with $1024^{3}$ gas elements and $1024^{3}$ dark matter particles evolved from $z= 249 \to 0$ assuming a \citet{Planck2016} concordant cosmology with $\Omega_{m}=0.3$, $\Omega_{\Lambda}=0.7$, $\Omega_{b}= 0.048$, $H_{0}=0.68\kmsmpc$, $\sigma_{8}=0.82$, and $n_{s}=0.97 $.  \simba\ accounts for radiative cooling and metagalactic photo-ionisation heating using the {\sc Grackle-3.1} package~\citep{Smith2017}, star formation based on a sub-grid $H_2$ model, stellar evolution assuming a \citet{Chabrier2003} initial mass function, chemical enrichment from Type~II and Type~Ia supernovae and stellar mass loss, and on-the-fly dust production and destruction~\citep{Li2019}.
Galactic winds are modeled kinetically following scalings from the FIRE simulations \citep{Muratov2015,DAA2017c}. 

Galaxies are identified using a friends-of-friends galaxy finder that is applied to all stars, black holes, and gas elements with a density above that of the star formation density threshold of $n_{H}>0.13$ H atoms~cm$^{-3}$. \simba\ stochastically spawns star particles from gas elements with the same mass. Galaxies are resolved down to 32 star particle masses which is equivalent to $\mstar=5.8\times10^{8}\msun$.

Galaxies that have reached a mass of $\mstar \geq \gamma_{\rm BH} \mbh$ with $\gamma_{\rm BH} = 3 \times 10^{5}$ are seeded with black holes with $\mbh = 10^{4} h^{-1}\msun $, based on the idea that small galaxies suppress black hole growth via stellar feedback~\citep{Dubois2015,Bower2017,DAA2017b,Habouzit2017}. This seeding occurs then for galaxies with $\mstar \approx 10^{9.5} \msun$, so any galaxy with $\mstar$ less than this threshold that hosts a black hole must have undergone some mass loss event.  In this work we will generally be considering these more massive galaxies that host radio jets. 

HDF5 catalogues of the pre-computed properties of galaxies at each snapshot are generated using the publicly available {\sc yt}-based package {\sc Caesar}\footnote{\tt https://caesar.readthedocs.io}. This includes both the black hole and dark matter halo properties of the galaxy, as well as the particle lists for these objects so that additional properties can be computed by the user. In this work we will be focused mostly on $z=0$ but when referring to evolution will make use of the snapshots at $z=3 \to 0$. Next we detail the sub-resolution models implemented to describe black hole accretion and feedback within \simba.

\subsection{Black Hole Accretion}
\label{sec:accretion}

\simba\ employs an accretion model for black hole growth based on two modes; the gravitational torque-limited accretion model~\citep{DAA2017} for cold gas ($T<10^{5}K$) and Bondi accretion~\citep{BondiAndHoyle,Bondi1952,Hoyle1939} from hot gas ($T>10^{5}K$), detailed as follows:

\subsubsection{Gravitational Torque-Limited Model}
\label{sec:gt}
Accretion rates for cold gas are based on the gravitational torque model of \citet{HopkinsQuataert2011} which estimates the gas inflow rate, $\dot{\rm M}_{\rm Torque}$, driven by gravitational instabilities from galactic scales down to the accretion disk surrounding the black hole as

\begin{equation}
\label{eq:GT2}
\begin{split}
\dot{\rm M}_{\rm Torque}\approx \epsilon_{\rm T}f^{5/2}_{\rm d} \times \Big(\frac{\mbh}{10^{8} \rm \msun}\Big)^{1/6}\Big(\frac{\mgas(\ro)+\mstar(\ro)}{10^{9} \rm \msun}\Big)\\
\times \Big(\frac{\ro}{100 \rm pc}\Big)^{-3/2}\Big(1+\frac{f_{0}}{f_{\rm gas}}\Big)^{-1} \rm \msun yr^{-1}
\end{split}
\end{equation}
where $f_{\rm d}$ is the disk fraction for the combined gas and stellar disk mass ${\rm M}_{\rm d}(\ro)$ within a distance $\ro$ such that:
\begin{equation}
\label{eq:GT3}
f_{\rm d}\equiv \frac{{\rm M}_{\rm d}(\ro)}{\mgas(\ro)+\mstar(\ro)}
\end{equation}
while ${\rm M}_{\rm gas}(\ro)$ and $\mstar(\ro)$ represent the total gas and stellar masses within $\ro$. $f_{\rm gas}$ is the gas fraction relative to the disk mass
\begin{equation}
\label{eq:GT4}
f_{\rm gas}\equiv \frac{\mgas(\ro)}{{\rm M}_{\rm d}(\ro)}
\end{equation}
and 
\begin{equation}
\label{eq:GT5}
f_{\rm 0}\approx 0.31 f^{2}_{\rm d}\Big(\frac{{\rm M}_{\rm d}(\ro)}{10^9 \msun}\Big)^{-1/3}.
\end{equation}
We define $\epsilon_{\rm T} \equiv \epsilon_{\rm m} \times \alpha_{\rm T}$, where $\alpha_{\rm T} = 5$ is the normalization of $\dot{\rm M}_{\rm Torque}$ proposed by \citet{HopkinsQuataert2011} and $\epsilon_{\rm m}$
is a free parameter introduced to account for processes that affect the radial transport of gas at unresolved scales. 
We tune this to $\epsilon_{m}=0.1$ in order to match the amplitude of the $\mbh-\mstar$ relation at $z=0$ \citep{DAA2017}.  $\ro$ is taken to be the radius enclosing 256 gas elements, with an upper limit of 2$\hkpc$. Evaluating equation \ref{eq:GT2} requires the separation of spheroidal and disk components within $\ro$, which is done by means of a kinematic decomposition \citep{DAA2013,DAA2015}.  Crucially, \citet{HopkinsQuataert2011} demonstrated using parsec-scale isolated disk simulations that this model yields consistent results even for $\ro$ as large as a kpc, which is comparable to the resolution in \simba.  This suggests that although far from able to resolve the relevant disk instabilities, \simba\ will still broadly capture the behaviour of torque-limited black hole accretion.

\subsubsection{Bondi-Hoyle-Lyttleton Parameterisation}
\label{sec:bondi}

The Bondi model is a prescription for black hole growth widely used in galaxy formation simulations \citep[e.g.][]{Springel2005b,Dubois2012,Choi2012}. The Bondi model states that for a black hole with mass $\mbh$ moving at a velocity $v$ relative to a uniform distribution of gas with density $\rho$ and sound speed $c_{s}$, the accretion rate of the black hole due to Bondi is given by
\begin{equation}
\dot{\rm M}_{\rm Bondi}= \alpha \frac{4\pi G^{2}\mbh^{2}\rho}{(c_{s}+v)^{3/2}}
\end{equation}
where $\alpha$ is a dimensionless parameter usually used to boost accretion rates and compensate partially for high mean gas temperatures as a consequence of the multi-phase sub-grid model of star formation and/or the lack of resolution required to resolve the Bondi radius. In \simba, we instead suppress $\dot{\rm M}_{\rm Bondi}$ by $\alpha \equiv \epsilon_{\rm m}=0.1$, which is the same factor we use in the $\dot{\rm M}_{\rm Torque}$ model.  For comparison to a real black hole, the Milky Way's black hole is accreting at around the Bondi rate, but the advection-dominated accretion flow (ADAF) rate is somewhat less~\citep{Quataert1999}. In \simba, we aim to be modeling this ADAF rate via Bondi accretion from hot gas.

\subsubsection{Numerical Implementation}

We apply the torque-limited accretion formula to all the gas within $R_0$ that has a temperature T$<10^5$K, while for T$>10^5$K gas we employ the Bondi formula, computing $\rho$ and $c_s$ from the hot gas only within $\ro$. A given black hole can thus accrete gas in both Bondi and torque-limited modes at any given timestep.  The total accretion onto the black hole is then
\begin{equation}
\label{eq:totaccr}
\dot{\rm M}_{\rm BH}=(1-\eta)(\dot{\rm M}_{\rm Bondi}+\dot{\rm M}_{\rm Torque})
\end{equation}
where $\eta=0.1$ is the radiative efficiency of the black hole. We limit Bondi accretion to the Eddington rate, while torque-limited accretion is capped at $3\times$ the Eddington rate owing to its lack of spherical symmetry. The time step of black hole particles is limited so that they do not grow $>0.1\%$ of their current mass in a single simulation time step to avoid large stochastic fluctuations, but this limit is very rarely invoked.

Black hole accretion proceeds stochastically \citep{Springel2005}. Gas particles within $\ro$ get a fraction of their mass subtracted and added to the black hole, with a probability that statistically satisfies the mass growth (equation \ref{eq:totaccr}).  If a particle is sufficiently small compared to its original mass, it is swallowed completely.

\subsection{Black Hole Feedback}
\label{sec:feedback}

Motivated by the observed dichotomy of accretion rates of AGN~\citep{HeckmanAndBest2014} and their corresponding outflows, \simba\ employs a multi-mode feedback model governed by the instantaneous Eddington ratio of the black hole. 
\subsubsection{Kinetic Feedback}
There is significant evidence suggesting that radiatively efficient AGN can drive strong winds~\citep{Sturm2011,Fabian2012}. For high $\fedd$ mode outflows, the outflow velocity of these winds can be estimated by the ionised gas linewidths of X-ray detected AGN from SDSS observations \citep{Perna2017a} and parameterised in terms of the black hole mass such that
\begin{equation}
    v_{\rm w,EL} = 500+500(\log\ \mbh -6)/3 \kms.
\end{equation}
These will be referred to as AGN winds.

If $\fedd<0.2$, we slowly begin transitioning to the jet mode where the velocity becomes increasingly higher as $\fedd$ drops:
\begin{equation}
    v_{\rm w} = v_{\rm w,EL}+7000\log_{10}\Big(\frac{0.2}{f_{\rm Edd}} \Big) \kms,
\end{equation}
with a cap to the velocity increase at 7000$\kms$ and resulting in a maximum jet speed at $\fedd\leq0.02$. Additionally, a criterion requiring $\mbh>{\rm M}_{\rm BH,lim}$ is added and motivated by observations that show that jets only arise in galaxies with velocity dispersions corresponding to black holes with $\mbh\gtrsim 10^{8} \msun$~\citep{Barisic2017,Mclure2004}. We conservatively choose ${\rm M}_{\rm BH,lim}=10^{7.5} \msun$. This mass limit prevents small black holes with temporarily low accretion rates from driving high powered jets.

AGN-driven outflows are modelled by stochastically kicking particles around the black holes with velocity $v_w$ with probability 
\begin{equation}
\label{eq:FB1}
p_{j} = \frac{1-f_{\rm m}}{f_{\rm m}} \times \frac{w_{\rm j}}{m_{\rm j}} \times \dot{\rm M}_{\rm BH} \Delta t
\end{equation}
where $w_{\rm j}$ is a kernel weight and $f_{\rm m}$ is the fraction of mass accreted by the black hole and subtracted from the gas particle before ejection~\citep{DAA2017}. This gives an outflow mass loading factor of $\dot{\rm M}_{\rm out}/\dot{\rm M}_{\rm BH} = (1-f_{\rm m})/f_{\rm m}$.  We set $f_{\rm m}$ for each outflow event such that the momentum ejected by the black hole is $20{\rm L}/c$, where L$=\eta \dot{\rm M}_{\rm BH} c^2$. \edit{The momentum choice} is based on the inferred energy and momentum inputs from \edit{observations of AGN outflows}~\citep{Fiore2017,Ishibashi2018}, albeit towards the upper end of the observations, \edit{which we found is required in \simba} in order to enact sufficient feedback in \simba\ to quench galaxies.

We note that the jets modeled in \simba\ (as with most cosmological-scale jet feedback models) represent so-called {\it heavy} jets, at relatively low velocities and high mass loading. Light jets that produce large lobes and FRII-like morphologies are thus not included per se. Thus the jet feedback in \simba\ should be regarded as a proxy for delivering a certain amount of energy and momentum on halo scales, although the details of this may be incorrect.  It is beyond the resolution of cosmological-scale simulations  to model the details of light jet launching or propagation.

\subsubsection{X-ray Feedback}

\simba\ additionally includes high-energy photon pressure feedback, although this does not play a significant role in the present work.  The energy input rate due to X-rays emitted by the accretion disk is computed following \citet{Choi2012}, assuming a radiative efficiency $\eta=0.1$.  In gas-rich galaxies, severe radiative losses are expected in the ISM, hence we only apply X-ray feedback below a galaxy gas fraction threshold of $\fgas<0.2$, and in galaxies with full velocity jets ($v_{\rm w}\ga 7000\kms$). The feedback is applied spherically within $\ro$, providing an outwards kick for star-forming gas and heating for non-starforming gas. This mode is important in understanding the hot gas in galaxy groups and clusters~\citep{Robson2020} and for providing a final evacuation of gas to fully quench galaxies~\citep{Dave2019}, which for instance manifests in green valley galaxy profiles and qualitatively improves agreement with observations~\citep{Appleby2019}.

\subsection{Modeling Radio Emission}
\label{sec:radio_em}

We aim to compute the 1.4~GHz radio emission for all resolved galaxies in \simba. We consider radio emission both due to star formation and accretion by the central SMBH.
Within the AGN population, we define HERGs and LERGs as those SMBHs with accretion rates dominated by gravitational torque limited accretion and Bondi accretion respectively.  We describe these aspects in more detail below.

\subsubsection{Star formation radio luminosity}
\label{sec:SF}

Radio observations trace the star formation within galaxies due to the synchrotron radiation emitted by relativistic electrons which arise from supernova and supernova remnants, as well as thermal bremsstrahlung from hot electrons in HII regions which is indicative of young hot stars.

We model the radio emission from star formation following \citet{Condon1992}, which depends on the star formation rate from a \citet{Chabrier2003} initial mass function (IMF) for stars more massive than $5 \msun$. The thermal and non-thermal contributions to a given galaxy's luminosity are computed as 
\begin{equation}
    \Big(\frac{{\rm P}_{\rm non-thermal}}{\rm W\ Hz^{-1}}\Big) = 5.3\times 10^{21} \Big(\frac{\nu}{\rm GHz}\Big)^{-0.8} \Big(\frac{\rm SFR[M \geq 5\msun]}{\msun {\rm yr^{-1}}}\Big)
\end{equation}

\begin{equation}
    \Big(\frac{{\rm P}_{\rm thermal}}{\rm W\ Hz^{-1}}\Big) = 5.5\times 10^{20} \Big(\frac{\nu}{\rm GHz}\Big)^{-0.1} \Big(\frac{\rm SFR[M \geq 5\msun]}{\msun \rm yr^{-1}}\Big)
\end{equation}

where $\nu$ is the observed frequency, in this case 1.4 GHz, and SFR$[{\rm M}\geq5\msun]$ is the rate of formation of stars larger than $5\msun$. 
The final luminosity is then the sum of the thermal and non-thermal luminosities.

\subsubsection{AGN radio luminosity}
\label{sec:RLAGN}
RLAGN are typically identified by the presence of relativistic jets that span scales from pc to Mpc scales in extreme cases, and are detected via synchrotron emission produced at radio wavelengths.  
LERGs emit most of their energy by means of these powerful jets, whereas HERGS radiate efficiently across the electromagnetic spectrum. However, a small percentage (roughly $10\%$) of HERGs exhibit incredibly powerful jets that emit strongly in the radio. This population is typically referred to as Radio Loud Quasars (RLQSOs) and narrow-line radio galaxies.

We identify galaxies in \simba\ that produce jets based on the criterion that they have $\mbh>10^{8} \msun$ and $0<\fedd<0.02$, i.e. where \simba\ black holes are accreting and jets are at full power, and consider these as comprising the radio loud AGN population. We also consider only galaxies that have not undergone any interactions that would trigger mass-loss to the point that the galaxy is lower mass than that of the mass at which it is seeded, as well as only SMBHs that have reached the $\mbh-\mstar$ relation in ~\citet{Thomas2019}. Both the black hole seeding and convergence onto the $\mbh-\mstar$ relation occur at $\mstar \approx 10^{9.5}\msun$.
 
To estimate the radio emission detected from these AGN, we sum the contributions from core and extended emission.  In each case, we apply empirical relations from the literature to connect the black hole accretion rate to the 1.4~GHz radio luminosity.  In practice, ``core'' is defined as radio emission on scales that are unresolved in the observations used to calibrate the relations we employ, in particular the FIRST radio survey with $\approx 5"$ resolution which is larger than the optical radius of the vast majority of the host radio-AGN host galaxies.  \simba's modest simulation volume does not include the rarest and brightest AGN population that are dominated by bright radio lobes such as Fanaroff-Riley Type II (FRII) objects. Instead the vast majority of our sources are expected to be core-dominated. As such, in \simba, we expect most AGN emission to be core-dominated.

For core emission, following \citet{Slyz2015} we employ the empirical relation from \citet{Kording2008}:
\begin{equation}
    \label{eq:rad_all}
    \frac{{\rm P}_{\rm Rad}}{10^{30}\rm erg\ s^{-1}} = \Big( \frac{\mdot }{4\times10^{17}\rm g\ s^{-1}} \Big)^{\frac{17}{12}},
\end{equation}
where ${\rm P}_{\rm Rad}\sim \nu {\rm P}_{\nu}$. In detail, this formula describes \edit{optically-thick} emission from the accretion disk region \edit{that is $\la$10~pc in size. Here we assume that, within the $\sim$kpc scales resolvable in \simba, the core emission is dominated by the optically-thick innermost region, while the emission from scales of tens of pc up to resolvable kpc scales is sub-dominant}.

We also include a separate contribution from extended emission owing to radio lobes, for a subset of the sources.
A troublesome point for \simba\ is that such lobe-dominated objects often have high accretion rates, but \simba\ produces essentially no massive black holes at $z=0$ with $\fedd\gg 0.02$.  This could be a failure of our model, or else simply a reflection of the rarity of such objects.

To include the RLQSO contribution, we provide an additional contribution to 10$\%$ of HERGs in the largest dark matter halos, while also meeting the criterion that they are the central galaxy in their halo, as given by

\begin{equation}
    \label{eq:rad_hergs}
    \log\ {\rm P}_{151} \big({\rm W\ Hz^{-1} sr^{-1}}\big) = \log\ \mdot  +0.15
\end{equation}
with ${\rm P}_{\nu}\propto \nu^{-0.7}$ to convert to $\power$. We will demonstrate in Appendix~\ref{AA} that the RLQSO population generally makes no contribution except for a minor one at the bright end, and other reasonable choices for populating RLQSOs in our simulation makes no significant difference to our results.

\subsubsection{HERGS $\&$ LERGS}
\label{sec:herglerg}

\begin{figure}
\begin{center}
\includegraphics[width=0.85\linewidth]{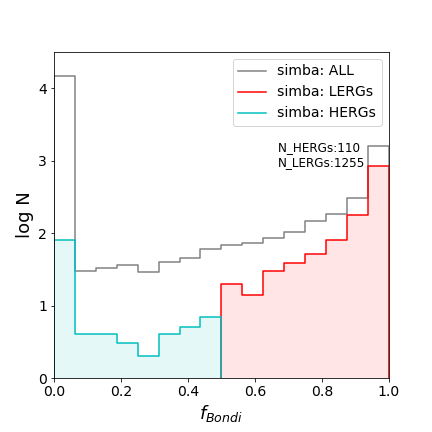}
\captionof{figure}{Distribution of the fraction of accretion rates in Bondi i.e $\frac{\dot{\rm M}_{\rm Bondi}}{\mdot}$ at z=0. The grey histogram shows the distribution for all SMBHs in \simba\ while the cyan and red histograms show the distribution for selected HERGs and LERGs respectively.}
\label{fig:fbondi_dist}
\end{center}
\end{figure}

Within the RLAGN population, we define HERGs and LERGs by their dominant mode of black hole accretion.  HERGs are assumed to occur when accretion is dominated by gravitational torque limited accretion, while LERGs occur when it is dominated by Bondi accretion.  Specifically, we compute $f_{\rm Bondi}=\frac{\dot{\rm M}_{\rm Bondi}}{\dot{\rm M}_{\rm BH}}$, and use $f_{\rm Bondi}=0.5$ as our threshold to separate these populations.
Because our accretion is implemented in a stochastic manner, we cannot meaningfully use the instantaneous accretion rate, and instead we must smooth over some timescale.
A reasonable choice is the dynamical time of the inner disk from which our black holes accrete, $\tau_{\rm dyn} \approx \sqrt{\frac{h^{3}}{G \mbh}}$, where we take $h$ to be half the smoothing length of the black hole kernel. We found that typical accreting black holes with $\mbh > 10^8 \msun$ have a median $\tau_{\rm dyn} \approx$ 60~Myr, so to smooth over numerical stochasticity, we take the average accretion rate over the past 50~Myr. Using this accretion rate, we calculate $f_{\rm Bondi}$. To check if this is a feasible timescale, we also tried 100~Myr, and this showed no substantive differences to our results.

Figure~\ref{fig:fbondi_dist}, grey histogram, shows the distribution of $f_{\rm Bondi}$ for all SMBHs in \simba\ at $z=0$.  The red and blue lines and histograms show the contributions specifically identified as RLAGN, separated into HERGs and LERGs. Note that the $y$-axis is a log scale.  As noted in the upper right, we identify a total of 1365 RLAGN at $z=0$, of which 1255 are LERGs and 110 HERGs.  This ratio is comparable to that seen in observations which show $\sim$1-2 orders of magnitude more LERGs than HERGs~\citep{BestandHeckman2012,Whittam2018,Williams2018}.

We see a significantly bimodal $f_{\rm Bondi}$ distribution. In other words, the majority of SMBHs accrete essentially entirely by gravitational torque limited accretion or entirely by Bondi. The sum of the RLAGN between the two end bins makes up $\sim32\%$ of the total RLAGN population which means that $\sim68\%$ of RLAGN live in the most extreme bins. We differentiate HERGs and LERGs at $f_{\rm Bondi}=0.5$, but reasonable changes to this divide makes no significant impact in any of our other results.  For instance, the number of RLAGN between $f_{\rm Bondi}=0.5-0.6$ only makes up $2\%$ of the RLAGN population.

\begin{figure}
\begin{center}
\includegraphics[width=0.85\linewidth]{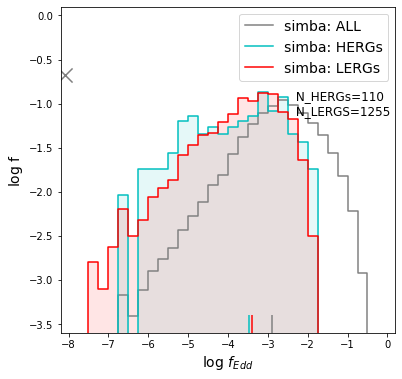}
\captionof{figure}{Fractional distribution of $\fedd$ for HERGs (cyan) and LERGs (red) compared to all galaxies with \edit{$\mbh>10^{6}$} (grey) at z=0. The solid lines at the bottom of the plot show the medians of each distribution. The grey cross shows the fraction of black holes with $\fedd=0$.}
\label{fig:fedds}
\end{center}
\end{figure}

Figure~\ref{fig:fedds} shows the fractional distributions of the Eddington fractions, $\fedd$, of HERGs and LERGs identified in \simba\ at z=0 and shown by the cyan and red shaded regions respectively. Although there are significantly fewer HERGs than LERGs produced by \simba , their distributions of $\fedd$ span a similar range with similar medians of $\log \fedd = -3.48$ and $-3.4$ respectively indicated by the solid cyan and red lines at the bottom of the panel. The condition employed by \simba\ that jet strength increases inversely with $\fedd$ already implies that RLAGN will have low accretion efficiencies and is amplified by the observed criteria of \edit{$\mstar>10^{9.5}\msun$} and \edit{$\mbh>10^{8}\msun$} which is where galaxies typically have less cold gas and star formation, and thus decreases the efficiency of accretion onto the black hole.
Observations show that HERGs have higher $\fedd$~\citep{BestandHeckman2012,HeckmanAndBest2014} while~\citet{Whittam2018}, using deeper observations, show the same trend though with significant overlap between the two populations. This might imply that with more sensitive radio observations, the observed dichotomy between HERGs and LERGs may be significantly reduced, which we will demonstrate in more detail later.

\section{RLAGN Demographics at Redshift 0}
\label{sec:RGs}

With the above framework, we can make predictions for the radio luminosity of \simba\ galaxies, separate them into star formation and AGN contributions, and classify them as HERGs or LERGs. We now examine the RLAGN luminosity function in these various categories, its evolution with redshift, and the relationship between RLAGN and the overall galaxy population in \simba. We begin by examining today's RLAGN population as predicted by \simba, describing the 1.4~GHz luminosity from HERGs and LERGs as the sum of their star forming and AGN contributions.
 
\subsection{Radio luminosity functions by galaxy type}

\begin{figure*}
\begin{center}
\includegraphics[width=0.8\linewidth]{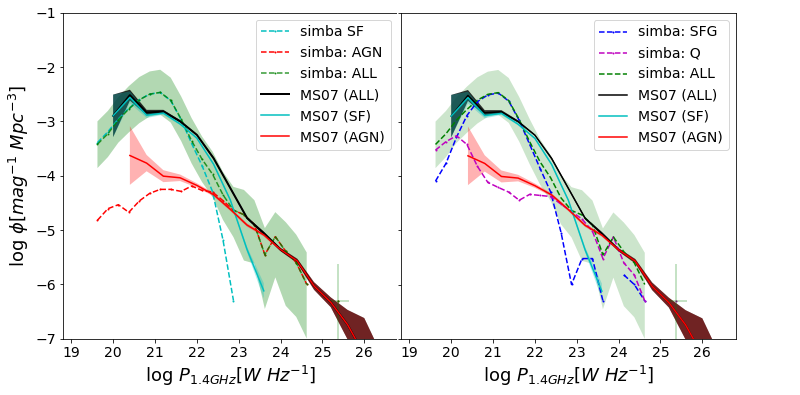}
\captionof{figure}{The RLFs of galaxies in the \simba\ simulations at $z=0$.  The left panel shows the contributions broken up by type of emission (from star formation or AGN), while the right panel shows it decomposed by star-forming versus quenched galaxies.  In both panels, the overall RLF is shown as the green line, with green shading representing estimate cosmic variance uncertainties.  Note that there is one individual galaxy shown at high radio luminosities; this is both an AGN-dominated and quenched galaxy.  The cyan/blue dashed line shows the contribution to the RLF due to star formation from all galaxies in the left panel, and due to star-forming galaxies in the right panel respectively.  The red/magenta dashed line in the left panel shows the radio AGN contribution from all galaxies that exhibit full velocity jets, while in the right panel it shows the RLF from quenched galaxies.  Observations from~\citet{MS2007} are shown by the black, cyan, and red lines corresponding to the total, star forming, and AGN radio luminosity functions respectively.  \simba\ does a good job of reproducing the overall RLF, as well as the contributions broken down by emission or galaxy type.}
\vskip-0.3in
\label{fig:RLF}
\end{center}
\end{figure*}

\begin{figure*}
\begin{center}
\includegraphics[width=0.9\linewidth]{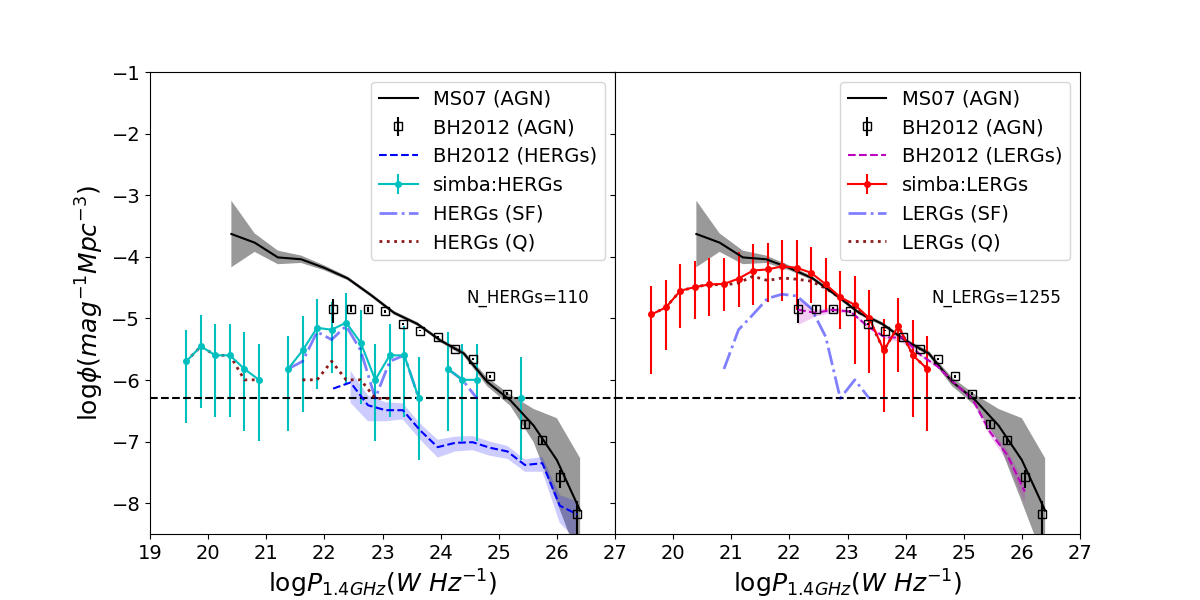}
\captionof{figure}{1.4 GHz RLF at z=0 for HERGs (cyan, left) and LERGs (red, right) in comparison with observations of the RLF of HERGs (blue dashed, left), LERGs (magenta dashed, right), and AGN RLF (black squares) by \citet{BestandHeckman2012}, as well as the AGN RLF by \citet{MS2007} (grey) for reference. Both HERGs and LERGs are split into star forming (blue dot-dashed lines) and quiescent (maroon dotted lines) populations.}
\label{fig:hergsVlergs}
\end{center}
\end{figure*}

Figure \ref{fig:RLF} shows the 1.4~GHz radio luminosity function (RLF) predicted by \simba, in comparison with observations. In both panels the green and black lines are the same and shows the total RLF from \simba\ and that observed by~\citet{MS2007}, respectively.  In the left and right panels, the \simba\ RLF is broken down in two different (albeit related) ways. In the left panel we determine separately the RLF contribution from star formation (cyan dashed line; \S\ref{sec:SF}) and black hole activity (red dashed line; \S\ref{sec:RLAGN}).  This is comparable to the separation used by \citet{MS2007}, so we colour the solid observational lines the same. In the right panel, we instead subdivide into the contribution to the RLF from quenched (Q) and star-forming galaxies (SFG), as magenta and blue dashed lines. We separate SF and Q galaxies based on an sSFR cut of $\frac{0.2}{t_{\rm H}}$~\citep{Pacifici2016,RodriguezMontero2019}.  In both panels, the green shading is an estimated cosmic variance from the variance over 8 sub-octants of the simulation volume.  For clarity the highest powered radio galaxy in \simba\ is plotted individually in green with its Poisson error.

The total RLF produced by \simba\ is a good match to the total observed RLF from \citet{MS2007}. At all radio powers, the observations are within the green shaded $1\sigma$ uncertainties for the simulation predictions.  This is an encouraging success for \simba, and indicates that \simba\ passes this basic test of the observed radio galaxy demographics.  That this was achieved without any ad hoc parameter choices tuned to match the RLF data is a more significant success, and fairly unique among current cosmological hydrodynamic simulations.

Looking at the left and right panels, we see that broadly, the breakdown in the RLF is similar whether one considers the separate contributions from star formation and AGN, or whether one breaks down the population into SF and Q galaxies.  In both cases, the AGN or quenched galaxy contribution begins to dominate at $\power\ga 10^{22.5}\WHz$. This suggests that the star formation dominates the radio power in star-forming galaxies, while the AGN dominate the radio power in quenched galaxies.  While this is not surprising, it didn't necessarily need to be the case if for instance there was substantial radio contributions from AGN in star-forming galaxies and vice versa.  In \simba, we don't expect this to happen because we define the radio contribution from AGN as arising from jets, which in \simba\ occur at low $\fedd$, and gas-rich galaxies that host star formation also tend to be growing black holes fairly rapidly and hence have high $\fedd$. However, we will see later (in Figure~\ref{fig:hergsVlergs}) that some AGN are hosted by star forming galaxies.

Comparing to the \citet{MS2007} observations now broken down by SFG and Q (right panel), we see that the \simba\ does a good job of separately reproducing these radio galaxy populations.  The quenched galaxies are in excellent agreement with the observed AGN RLF, while the star-forming galaxies are slightly under-predicted in radio power, but this is mostly within the $1\sigma$ uncertainties due to the low space density at $z=0$. We do not show the uncertainties on the SFG and AGN predictions individually to avoid clutter, but they are usually slightly larger in extent to the green shading. The observations, like \simba, show a transition from SF-dominated to Q-dominated galaxies at $\power\sim 10^{22-23}\WHz$, as well as a power-law behaviour towards the largest luminosities.  

In summary, \simba's model for black hole jets following \citet{BestandHeckman2012}, and their association with radio AGN, seems to produce radio galaxies both in the correct numbers as observed in the overall RLF, as well as associated with the correct types of galaxies.  There is a transition from SF-dominated radio emission in star-forming galaxies at $\power\la 10^{22.5}\WHz$, to AGN-dominated quenched galaxies above this.  While the SF contribution drops off exponentially at high luminosities, following the SFR function~\citep{Dave2019}, the AGN contribution continues as a power law to the highest luminosities.  \simba\ thus provides a plausible platform to examine the RLAGN population and its relationship to the overall galaxy population, as we do in the remainder of this paper.

\subsection{HERG and LERG luminosity functions}

\simba's unique two-mode black hole accretion model is based on the same underlying physical dichotomy believed to drive the separation between HERGs and LERGs~\citep{HeckmanAndBest2014}.  Hence in \simba, we have the ability to examine these populations independently, by associating torque-limited accretion from cool gas as ``HERG" accretion, and Bondi accretion from hot gas as ``LERG" accretion, as described in \S\ref{sec:herglerg}.  We remind the reader that in \simba\ the accretion is subdivided into these modes on kpc scales, while in reality it is occurring on far smaller accretion disk scales.  Nonetheless, this association represents a physically motivated ansatz that we explore here.

Figure~\ref{fig:hergsVlergs} shows, in the left and right panels, the RLF for HERGs (cyan points with errorbars) and LERGs (red points with errorbars) respectively.  We compare to observations of HERGs and LERGs by \citet{BestandHeckman2012} (blue and magenta dashed lines) as well as showing their AGN RLF (black squares) and reproducing the \citet{MS2007} AGN RLF (black line) for reference. Furthermore, HERGs and LERGs are both split into SF (blue dot dashed) and Q (maroon dotted) host galaxy populations. The horizontal black dashed line shows the minimum space density representable by a single object within \simba\ owing to its 100$\hmpc$ volume.  Note that the $y-$axis in this plot goes to substantially lower values than in Figure~\ref{fig:RLF}, to show the continued power law in the RLF, but as is evident \simba\ is unable to probe these very low volume densities.  Hence we restrict our comparisons to the data above the horizontal dashed line.

Focusing on the LERG population (right panel), \simba\ yields excellent agreement with observations in the overlapping region.  There is a turnover at the lower radio powers that may hint at disagreement with observations, which forthcoming deeper surveys would confirm if true.  As we expect, the LERG host galaxies are predominantly quiescent, although there are a handful of galaxies that are star-forming; we will examine the host galaxy demographics in more detail in the next section.  This comparison further highlights the success that \simba\ has in reproducing radio AGN luminosities.

The HERG population, in contrast, is not quite in such good agreement, though the predictions are still generally within $1\sigma$ uncertainties of the data.  There is an overproduction of HERGs by $\sim\times 2-3$ at overlapping radio luminosities.  Indeed, the expectation from the data is that we would produce basically no HERGs at all within our \simba\ volume. Our limited volume is unable to produce the brightest sources observed in observations and are thus limited to what are typically described as low-luminosity sources. We note that the stochastic nature of cool gas accretion may result in a duty cycle for HERGs.  As a rough example, if we were to assume a duty cycle of $\sim0.2$ for HERGs, this would decrease the number density by $\sim\times 5$, while increasing the luminosity of each HERGs by a factor of $\sim\times 5$, resulting in a better match the observations. We will consider the impact of duty cycles that are below the temporal resolution of cosmological simulations in future work.  Still, it is encouraging that the shape of the HERG RLF is similar to that observed, being rather flat. We note that, at high luminosities, HERGs are dominated by star-forming galaxies in \simba.  We remind the reader that in our classification, green valley galaxies are counted as SF; the exact distribution in sSFR will be discussed in the next section.

In summary, \simba\ broadly reproduces the properties of both HERGs and LERGs separately, albeit with an apparent overproduction of HERGs.  In general, the LERG radio power is dominated by quenched galaxies, while the HERGs are mostly galaxies having some star formation.
\simba's general success in reproducing HERGs and LERGs is highly encouraging, particularly since it was achieved without any significant tuning in model parameters.  As noted in \S\ref{sec:herglerg}, it is also insensitive to the choice in \simba\ for classifying HERGs vs. LERGs.  While this does not definitively prove our ansatz relating HERGs and LERGs to accretion mode, it is suggestive that there is sufficient connection between gas properties at kpc scales and time-averaged properties on accretion disk scales to justify our ansatz.  This is, as far as we are aware, the first time a cosmological hydrodynamic simulation has been able to reproduce the properties of HERGs and LERGs. \edit{Not only will these results act as predictions for upcoming surveys, but these surveys will in turn serve as a test of our model.}

\section{Evolution of the RLAGN population}
\label{sec:Evolve}

RLAGN have been observed to very high redshift and beyond Cosmic Noon, when both the cosmic star formation rate density and black hole accretion rate density peaked~\citep[e.g.][]{MadauDickinson2014,Smolcic2017}.  Studying the redshift evolution of the RLAGN population thus provides a complementary test of \simba's modeling.  Using the same procedure for estimating the 1.4 GHz radio power at $z=0$ (\S\ref{sec:SF} and \S\ref{sec:RLAGN}), here we compute the RLFs from \simba\ for $z=3\to0.25$, and compare to observations.

\subsection{Radio luminosity function evolution}

\begin{figure*}
\begin{center}
\includegraphics[width=0.8\linewidth]{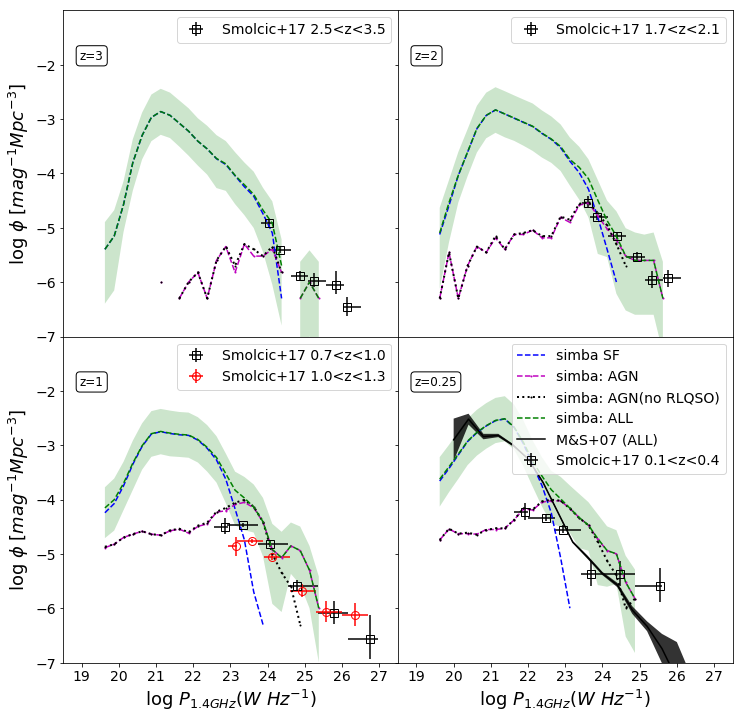}
 \captionof{figure}{Evolution of the RLF from $z=3\to0.25$. Blue dashed lines show the contribution of star formation to the RLF while magenta lines show that of the AGN and the total RLF shown as green dashed lines with green shading representing the cosmic variance uncertainties as in figure~\ref{fig:RLF}. The black dotted line shows the AGN RLF if the radio power from RLQSOs are not considered. At $z=0.25$ we compare the RLF to that of~\citet{MS2007} (black line) as well as observations by~\citet{Smolcic2017} represented by black squares. For $z=3\to1$ we compare to observations by~\citet{Smolcic2017} represented by black squares as well as red circles at $z=1$.}
\label{fig:RLFev}
\end{center}
\end{figure*}

Figure~\ref{fig:RLFev} shows the evolution of the RLF from $z=3\to 0.25$, analogous to what was shown for $z=0$ in Figure~\ref{fig:RLF}. The green dashed line shows the total RLF in \simba, with the green band representing cosmic variance uncertainties. The RLF is split into contributions from star formation processes (blue dashed) within all resolved galaxies, as well as contribution from AGN with ongoing jet feedback (magenta dashed). We compare to observations by~\citet{Smolcic2017} represented by empty black squares as well as empty red circles at $z=1$.  At $z=0.25$, we also reproduce the $z\sim 0$ \citet{MS2007} data (black line) for reference. In addition, we show the effect of disregarding the radio emission from RLQSOs (\S\ref{sec:RLAGN}) as the dotted black line, and note that the only effect this has is a slight decrease of the RLF at the highest luminosities, however this small difference is not significant to further results presented throughout this paper. 

\simba\ broadly predicts that the overall RLF is fairly similar in amplitude across all redshifts.  At $z\sim 3$, a small number of the brightest RLAGN ($\power \sim 10^{25}\WHz$) have appeared, but by $z\sim 2$ they increase and are in place persisting to $z\sim 0$.  Breaking this down into SF vs. AGN, the brightest galaxies are always AGN-powered, but the luminosity at which the transition from SF-powered to AGN-powered occurs increases by $\sim 0.5$~dex in $\power$ per unit redshift.  The SF contribution to the  RLF has a steep faint end slope, and the turnover at low $\power$ likely owes to resolution limits of the simulation and the assumed sharp cutoff in $\mstar$ and $\mbh$ for forming RLAGN, so should not be considered a strong prediction of the model.

Comparing \simba\ to the \citet{Smolcic2017} observations, we see quite good agreement at all redshifts.   Indeed, the worst agreement comes at low redshift, where \simba\ appears to mildly overpredict the observations.  We note one caveat, however:  This agreement at high-$z$ is somewhat sensitive to our choice of having RLAGN be those with $\fedd<0.02$.  Recall that in \simba, the jet velocity boost begins to kick in already at $\fedd<0.2$, although its logarithmic dependence on $\fedd$ means that it does not become very strong until much closer to $\fedd=0.02$.  We choose RLAGN as those having full-velocity jets, which makes little difference at $z=0$ because (as we discuss next) very few galaxies have $\fedd>0.02$, but at higher redshifts this is not the case. For instance, if we were to define RLAGN as those with $\fedd<0.2$ (in the other extreme), then the RLF at $z\sim 1-3$ would shift up by $\sim 0.5-1$~dex in $\power$, and the agreement with data would be somewhat poorer. We may also need to investigate a different coupling between the accretion rate and the jet production/radio luminosity. Nonetheless, we feel that RLAGN being associated with fully powered jets is the most natural definition.

Overall, \simba\ predicts a RLF evolution that is broadly in agreement with observations.  With deeper and more sensitive observations such as the MIGHTEE survey, we will be able to map out the fainter end of the RLF across a large range of redshifts. Probing below $\power\la 10^{23}\WHz$ will provide insights into the role that star formation and AGN play in the evolution of galaxies, and moreover provide an independent extinction-free avenue to quantify the cosmic evolution of the star formation rate density.

\subsection{Eddington fraction evolution}

\begin{figure}
\begin{center}
\includegraphics[width=0.52\textwidth]{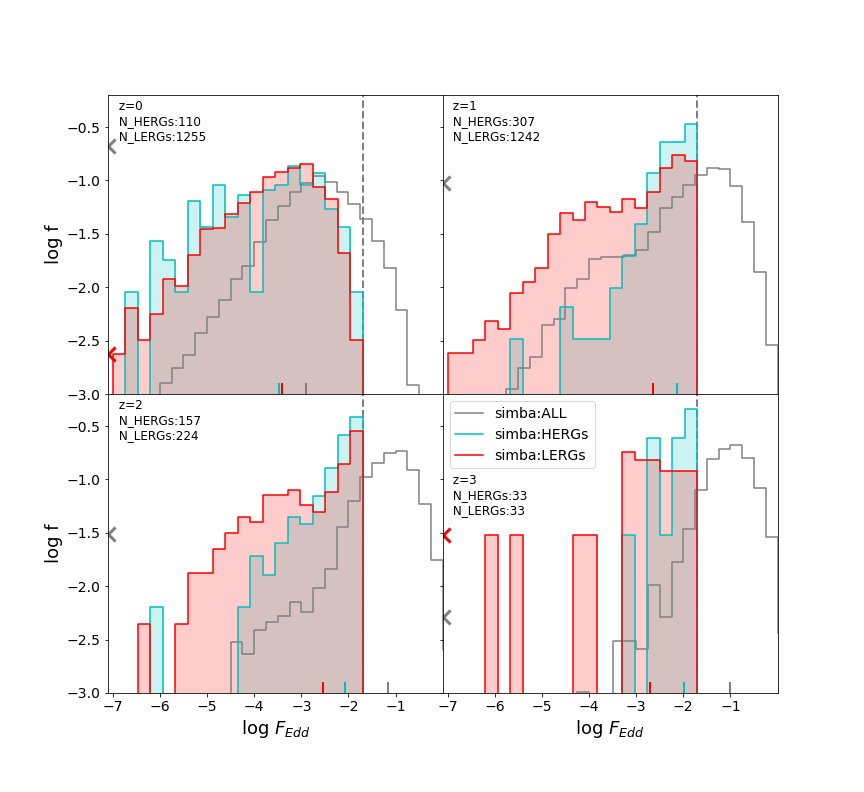}
\vskip-0.2in
\captionof{figure}{The evolution of Eddington fractions of HERGs (cyan) and LERGs (red) from $z=3\to0$. Grey lines show the fractional distribution of Eddington fractions for all galaxies. Grey and red crosses show the fraction of galaxies and LERGs with $\fedd<10^{-7}$ respectively.}
\label{fig:Fedd_Ev}
\end{center}
\end{figure}

\citet{DAA2013,DAA2015} showed that the torque-limited accretion model naturally yielded typical $\fedd$ values that dropped with time.  This is because at earlier epochs the galaxy accretion rates are higher, which translates into higher torque-limited accretion from the inner disk, and black holes are smaller.  It is thus interesting to quantify this in more detail within a full cosmological context, here focusing on the RLAGN population.  Besides elucidating some physical trends, this provides predictions for upcoming surveys that aim to quantify both accretion rates and black hole masses out to high redshifts.

Figure~\ref{fig:Fedd_Ev} shows the evolution of the Eddington fractions of HERGs (cyan shading), LERGs (red shading), and all galaxies with black holes of masses \edit{$\mbh >10^{6}\msun$} (grey line) from $z=3\to0$. The red and grey crosses shows the number of LERGs and all galaxies with $\log\fedd < -7$ respectively. The vertical dashed line represents our RLAGN (i.e. full jet feedback) threshold of $\fedd=0.02$.  The number of HERGs and LERGs are indicated in the upper left, which can be summed to give the total RLAGN counts within \simba's volume.

\simba\ predicts an overall $\fedd$ distribution that peaks at $\fedd\sim 10^{-3}$ at $z=0$, increasing steadily with redshift to $\fedd\sim 0.1$ at $z\sim 2-3$, in agreement with the post-processed simulations of \citet{DAA2015}.  The RLAGN by definition occupy the lower end this distribution, and so have typically lower $\fedd$.  At $z=0$, it is clear from the tailing distribution to high $\fedd$ that the imposed $\fedd$ cutoff does not strongly impact the RLAGN numbers.  However, even by $z\sim 1$, the RLAGN distribution is strongly truncated by the imposed cutoff.  While this cutoff is physically-motivated~\citep{HeckmanAndBest2014}, the predicted RLAGN counts become more sensitive to the exact threshold as noted in the previous section.  In \S\ref{sec:RLAGN} we define RLAGN as black holes hosting jets. As mentioned, jets are induced at $\fedd = 0.2$ with full velocity jets occurring at $\fedd<0.02$. An interesting observation at $z=0$ is that there is only one AGN between $\fedd = 0.02 - 0.2$, and thus no significant differences would be seen between choosing one limit over the other. However, as we move to higher redshifts, $\fedd$ for RLAGN increases (as well as that for all galaxies as seen in~\citet[figure 5 ]{Thomas2019}). With this, the number of RLAGN between 0.02 and 0.2 increases significantly. For higher redshifts, the selection of black holes with jets becomes a more important consideration. 

\simba\ shows a significant difference between the HERG and LERG populations at higher redshifts.  At $z=0$, there is not a significant difference between the two populations as indicated by a K-S test p-value$\sim 0.03$, and a similar median as mentioned in \S\ref{sec:herglerg}, but at higher redshifts the HERGs typically have higher $\fedd$ values.  As mentioned earlier, \citet{BestandHeckman2012, Whittam2018} shows that HERGs typically have higher $\fedd$ than LERGs, so it is interesting to find that \simba\ shows no obvious trend at $z=0$. It is especially interesting as in \simba, HERGs are chosen to be the cold/gravitational torque limited accretors which are associated with higher sSFRs as per figures~\ref{fig:med_props} and~\ref{fig:pfs} and would typically be seen at higher $\fedd$ values. However, the separation of RLAGN as those with high $\mstar$ and $\mbh$ limits the RLAGN population to the low end of $\fedd$. 

\section{RLAGN host properties}
\label{sec:host_props}
With a plausible framework in hand to model RLAGN, we can now take advantage of \simba's full galaxy formation model to investigate the nature of RLAGN and their relationship to the overall galaxy population.  

\subsection{The host galaxies of HERGs vs. LERGs}

\begin{figure*}
\begin{center}
\includegraphics[width=0.95\linewidth]{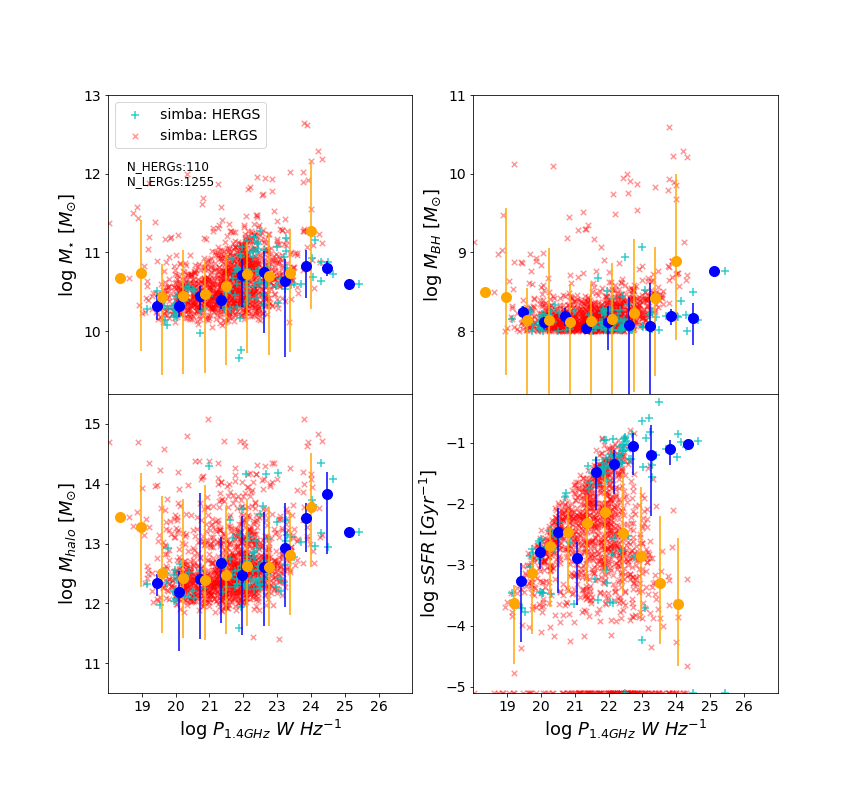}
\captionof{figure}{Global properties of HERGs (cyan plusses) and LERGs (red crosses) as a function of $\power$ at z=0. \textit{Top left to right}: $\mstar$, $\mbh$. \textit{Bottom left to right}: $\mhalo$, sSFR. Blue points with errorbars show the median value and 1$\sigma$ variance of the respective property in a given $\power$ bin for HERGs while the orange points show that of the LERGs. For the bottom right panel, the accumulation of points at \edit{log sSFR}= -5.1 shows all RLAGN with sSFR$< 10^{-5}$~Gyr$^{-1}$. }
\label{fig:med_props}
\end{center}
\end{figure*}

\begin{figure*}
\begin{center}
\includegraphics[width=0.9\linewidth]{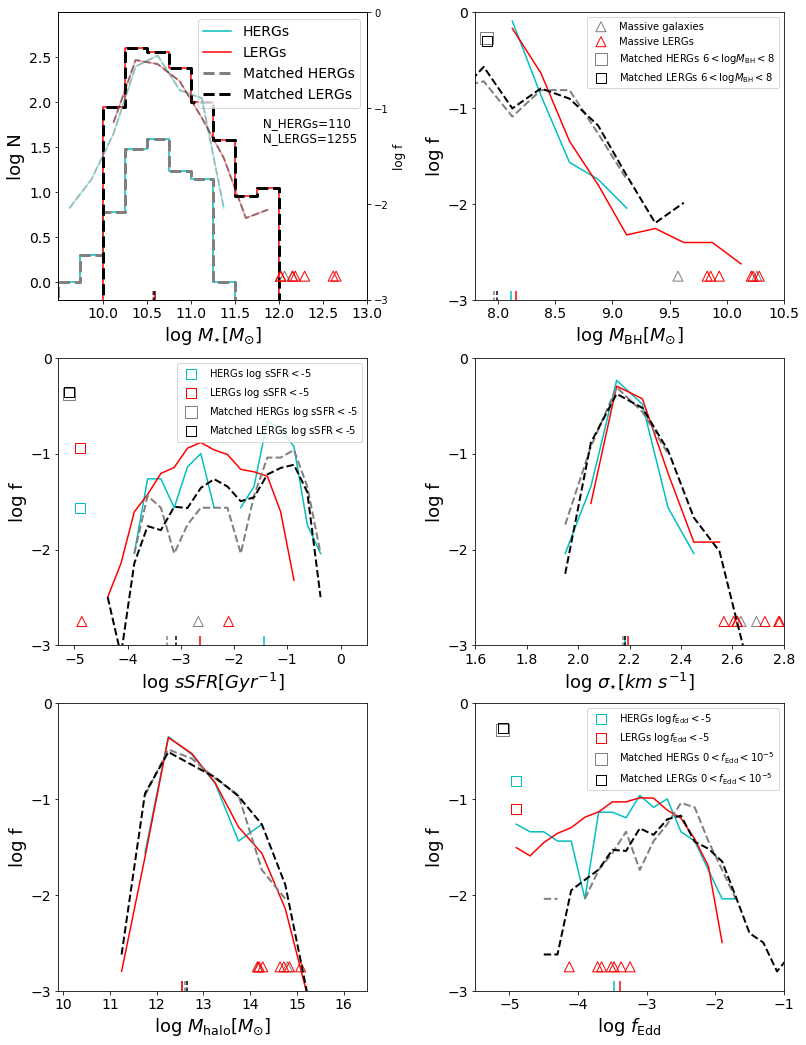}
\captionof{figure}{Distributions of the host galaxy properties at z=0 of HERGs (cyan) and LERGs (red) compared to the overall sample of galaxies in \simba\ that host a SMBH with $\mbh>10^{6} \msun$ and that are matched to the same stellar mass distribution of HERGs (grey dashed) and LERGs (black dashed).
\textit{Top left to right}: stellar mass $\mstar$; histogram shows the number distribution while the fainter lines show the fractional distribution, black hole mass $\mbh$.
\textit{Middle left to right}: specific star formation rate sSFR, velocity dispersion $\sigma_{\star}$. 
\textit{Bottom left to right:} halo mass $\mhalo$, Eddington fraction $\fedd$.
Squares show the fraction of each population with either $\mbh<10^{8} \msun$, \edit{sSFR$<10^{-5}$~Gyr$^{-1}$} or $\fedd <10^{-5}$. Triangles show the global properties of the 9 most massive galaxies in \simba. Solid and dashed lines show the median of the distributions for RLAGN and mass-matched samples respectively.}
\label{fig:pfs}
\end{center}
\end{figure*}

An interesting question to confront is whether at a particular radio luminosity, we expect to see differences between the host properties of HERGs and LERGs, and whether these properties change as a function of radio power.  Recall that in our model, HERGs and LERGs are classified based on whether they are dominated by torque-limited (cold mode) or Bondi (hot mode) accretion in the past 50~Myr (respectively).  Canonically, LERGs are thought to be found mainly in massive quiescent galaxies, while HERGs are usually regarded as occupying more moderate-mass star-forming galaxies. We investigate this here, in terms of the host galaxies' stellar mass $\mstar$, black hole mass $\mbh$, halo mass $\mhalo$, and specific star formation rate sSFR.  

Figure~\ref{fig:med_props} shows the host galaxy properties of HERGs and LERGs as a function their radio power. Cyan pluses are HERGs while red \edit{crosses} are LERGs, and blue and orange points with errorbars are the running medians in a given luminosity bin for HERGs and LERGs respectively. Clockwise from top left is the stellar mass $\mstar$, black hole mass $\mbh$, sSFR, and halo mass $\mhalo$ of the host galaxy, all as a function of 1.4~GHz radio power $\power$.

For the bulk of the population across most radio powers, \simba\ predicts little separation between the typical host galaxy masses of HERGs and LERGs. All these galaxies tend to live in moderate stellar mass galaxies near the knee in the stellar mass function, have moderate-mass black holes of $\mbh\sim 10^{8-9}\msun$, and live in group-sized halos of $\mhalo \approx 10^{12-14} \msun$.  This is perhaps surprising, since at face value it contradicts the conventional view of a dichotomy in the HERG and LERG host galaxy population. \edit{However, we note that \simba's radio galaxies are shifted towards less massive hosts than observations such as \citet{BestandHeckman2012}, whose radio galaxy sample spans primarily between $10.5<(\mstar/\msun)<12$.}

Closer inspection, however, begins to elucidate some differences.  These are most evident in the sSFR plot, where for all but the lowest radio powers, the HERGs clearly are living in more star-forming systems, while LERGs are mostly in quiescent galaxies with sSFR$<10^{-2}$~Gyr$^{-1}$. Hence if one focuses on the brighter RLAGN \edit{which tend to live in more massive galaxies,} we recover the canonical trend.  That said, \simba\ predicts that deeper surveys will find that HERG and LERG hosts have more similar star formation properties.

Another clear difference emerges when examining only the largest galaxies. For instance, all galaxies with $\mstar >2\times 10^{11}\msun$ are LERGs, and likewise all $\mbh> 10^{9}\msun$ are LERGs.  Hence the most massive galaxies with the largest black holes indeed do have strong preference for hosting LERGs, as seen observationally.  Interestingly, such a scenario does not play out the same way in halo mass -- even the largest halos are just as likely to host LERGs or HERGs.  Hence the HERG/LERG dichotomy in massive systems is more reflected in the black hole and host stellar mass, rather than host halo mass.

In summary, LERGs are the dominant population in the most massive galaxies with the largest black holes.  But aside from those rare objects, the bulk of the HERG and LERG populations reside in similar host galaxies, near the knee of the stellar mass function.  The one clear difference is that, at $\power \ga 10^{22}\WHz$, HERGs live in typically more star-forming galaxies, while LERGs are confined to quiescent galaxies.  At low radio power, the two populations blend together even in sSFR.  Hence \simba\ predicts that deeper radio surveys will find the host galaxy properties of HERGs and LERGs to be increasingly similar.

\subsection{HERGs and LERGs vs. the overall galaxy population}

RLAGN represent a minor fraction of the overall galaxy population. It is thus interesting to ask whether these galaxies preferentially live in certain types of hosts, from among the total population of galaxies with black holes. To study this, here we examine the relative fractions of HERGs and LERGs vs. the overall galaxy population as a function of the four galaxy properties we used the previous section, $\mstar$, $\mbh$, $\mhalo$, and sSFR, as well as velocity dispersion $\sigma_{\star}$ as a measure of morphology, and the accretion efficiency $\fedd$. From the overall galaxy population with $\mbh>10^6 \msun$ we create a stellar mass-matched sample to or HERG and LERG host galaxies, so as to remove any dependence of the results on differences in $\mstar$. RLAGN make up ~6.7$\%$ of these galaxies, but 7/9 of the most massive galaxies ($\mstar>10^{12} \msun$) are also RLAGN so we choose to exclude them from the overall galaxy population in order to more robustly compare the properties of galaxies, and instead separately discuss the results for these 9 galaxies.

Figure~\ref{fig:pfs} shows the fractional distributions of host galaxy properties of HERGs (cyan curves) and LERGs (red), vs. the mass-matched HERG and LERG samples (grey and black dashed lines, respectively) as described above, at $z=0$. The top row shows the stellar mass $\mstar$ which also shows the number distribution as a histogram, and black hole mass $\mbh$. The middle row shows the sSFR and velocity dispersion $\sigma_{\star}$, while the bottom row shows the halo mass $\mhalo$ and $\fedd$. 
Since the overall populations are chosen to have $\mbh>10^6 \msun$, we show the fraction of galaxies in the mass-matched sample with \edit{$10^{6} <(\mbh/\msun)<10^{8}$} as grey and black squares for HERGs or LERGs, respectively, in the $\mbh$ panel.
In the sSFR and $\fedd$ panels, we further show red and blue squares for the fraction of the LERG and HERG population with sSFR$<10^{-5}$~Gyr$^{-1}$ (typically with zero SF) and $\fedd<10^{-5}$ as well as that of the mass-matched sample in grey and black.
We show the median of each property as a solid (for RLAGN) or dashed (for mass-matched) tick mark at the bottom of each panel. For the 9 most massive galaxies, we show them individually as red or grey triangles defined by whether they are a LERG or normal galaxy. Note that there are no HERGs at the highest masses, and the position of the triangle depends only on the value of the x-axis, i.e. the amplitude/fractional value is of no significance. Also, note that these are fractional histograms within each given population (except for $\mstar$ which shows both the numbers and fraction), in order to see where HERGs and LERGs arise versus each other and within the overall population, for each property.

\subsubsection{HERGs vs LERGs}

We begin by comparing the host galaxy properties of the HERGs (cyan lines) vs. LERGs (red lines).  In the upper left plot showing the distributions in stellar mass, there is only a weak difference in $\mstar$: The typical LERG is shifted to slightly higher stellar masses, so that the largest $\mstar$ galaxies exclusively host LERGs, while the few RLAGN at $\mstar<10^{10}\msun$ are exclusively HERGs. 

In the upper right panel, once a RLAGN's black hole exceeds $\mbh\ga 10^{9}\msun$, it will almost invariably host a LERG. The most massive LERGs also happen to host the most massive black holes, as shown by the red triangles, which is a natural consequence of the $\mstar-\mbh$ relation~\citep[see Figure 2 ]{Thomas2019}.

There is less difference over much of the range of sSFR (middle left) as both HERGs and LERGs have a wide range of sSFR.  In detail however, at the highest sSFRs going into the green valley regime (sSFR$\ga 10^{-2}$~Gyr$^{-1}$), the population becomes increasingly dominated by HERGs, and LERGs are absent altogether in main sequence galaxies (sSFR$\ga 10^{-1}$~Gyr$^{-1}$). The sSFRs for the most massive LERGs are not resticted to any location, however 5 of the 7 most massive LERGs exhibit sSFR$\la 10^{-5}$Gyr$^{-1}$. Of the remaining 2, only one massive LERG exhibits large sSFR.

The velocity dispersion (middle right) and halo mass (lower left) distributions show even less difference between HERGs and LERGs. However, it is noted that the most massive LERGs are hosted at the highest $\sigma_{\star}$ and $\mhalo$.

There is also very little difference between the $\fedd$ distributions of HERGs and LERGs, with only a small fraction of HERGs appearing at $\fedd \ga 10^{-2}$ while at the same time hosting a larger fraction with $\fedd \la 10^{-5}$ than that of LERGs. In addition, the most massive galaxies lie more closely to the median of the distribution, rather at any particular extreme.

The overall trend from Figure~\ref{fig:pfs} is that HERGs and LERGs broadly live in similar types of galaxies, with minor differences in which LERGs reach higher stellar, black hole, and halo masses, higher velocity dispersions, as well as lower sSFRs.  The most distinct trend is that HERGs are increasingly more common compared to LERGs at high sSFR, with LERGs being completely absent in main sequence galaxies.  This general similarity as well as the specific difference were also qualitatively evident in Figure~\ref{fig:med_props}.

\subsubsection{RLAGN vs the overall galaxy population}

We now compare both the HERGs and LERGs (i.e. all RLAGN) to the overall galaxy population, matched in $\mstar$.  As seen in the upper left panel of Figure~\ref{fig:pfs}, the stellar mass distribution of the mass-matched samples are identical to the RLAGN populations by construction, but the remaining panels test how well LERGs and HERGs trace all galaxies with the same $\mstar$.  2 of the 9 most massive galaxies are ``normal'' galaxies and represented by grey triangles, but since this is such a small comparison sample, we cannot statistically compare these extreme systems.

To quantify the distributions of HERGs and LERGs, as well as to that of the mass-matched samples, we perform two-sample Kolmogorov-Smirnov (KS) tests and record the associated $p$-values in Table~\ref{table:Pvalues}. The first column shows the $p$-values for the comparisons of HERGs vs LERGs, while the second and third columns show the $p$-values for HERGs and LERGs vs their mass matched samples, respectively. Each row describes an associated property in Figure~\ref{fig:pfs}. 

The differences between the RLAGN and mass-matched samples are particularly dramatic in $\mbh$ (upper right panel).  Fractionally, the RLAGN are dominant at both low and high $\mbh$ compared to the mass-matched samples.  The most massive black holes $\mbh>10^9 \msun$ are typically hosted by LERGs, while intermediate masses \edit{ $10^{8.5} \la (\mbh/\msun) \la 10^{9.5}$} are dominated by normal galaxies.  Unsurprisingly, the K-S test (see Table~\ref{table:Pvalues}) shows that the matched HERG and LERG samples are quite different in their $\mbh$ distributions than the true HERGs and LERGs in \simba.

In sSFR (middle left panel), the mass-matched sample spans a wide range going from little to no star formation through the green valley and to the main sequence. The $\mstar$ range of RLAGN thus spans the range of stellar masses over which galaxies go from predominantly star-forming to quenched.  The distribution of HERGs is somewhat similar to the mass-matched HERG sample, albeit HERGs are absent at the very lowest sSFR.  The K-S $p$-value (Table~\ref{table:Pvalues}) shows a significant difference between the HERGs and mass-matched HERGs.  The LERGs are more obviously different than the mass-matched sample, as they are completely absent at high sSFR.

Examining the velocity dispersions in the middle right panel, the mass-matched LERGs (black dashed line) is shifted to slightly lower $\sigma_{\star}$ than that of the LERGs, which makes them statistically inconsistent with being drawn from the same distributions. The distributions are also slightly different for the HERGs, but in this case, because of the lower numbers, the K-S test shows that these distributions are indistinguishable.

Similar to the velocity dispersions, the halo masses (bottom left) of RLAGN span the same range of that of the mass-matched sample, with very similar medians. As with $\sigma_\star$, the distributions for the HERGs are deemed more similar than for LERGs, although in neither case are the differences between the true and mass-matched samples striking or systematic.

Finally, looking at the Eddington ratios (lower right), here we see clear differences.  RLAGN are shifted towards lower $\fedd$ values compared to the mass-matched sample.  Particularly, once $\fedd<10^{-3}$, RLAGN have significantly higher fractional distributions than the mass-matched samples.  This isn't simply because RLAGN are selected to have $\fedd<0.02$, since not many black holes at $z=0$ have higher $\fedd$ in any case.  However, we note that there is a large fraction of the mass-matched sample that hosts black holes that are not currently accreting (whereas RLAGN must have some nonzero accretion), which results in a very low median which is not within the range of the plot.

\begin{center}
\begin{tabular}{l c c c }
 & RLAGN & Matched HERGs & Matched LERGs \\
 \hline 
 $\mstar$ & 0.86789 &  0.98006 & 0.24842 \\
 $\mbh$ & 0.00369 & $< 10^{-5}$ & $< 10^{-5}$ \\
 sSFR & $< 10^{-5}$ & $< 10^{-5}$ & $< 10^{-5}$\\
 $\sigma_{\star}$ & 0.13666  & 0.03607 & $< 10^{-5}$\\
 $\mhalo$ & 0.91164 & 0.07501 & $< 10^{-5}$ \\
 $\fedd$ & 0.03286 & $< 10^{-5}$ & $< 10^{-5}$
\end{tabular}
\captionof{table}{Two-sample KS test p-values for probability functions of properties of RLAGN and their mass-matched samples.}
\label{table:Pvalues}
\end{center}

In summary, our results here reiterate and quantify the results from the previous section, that LERGs typically live in larger galaxies with higher black holes masses, but they have little difference in halo mass and sSFR except for a population of HERGs living in the green valley.  Meanwhile, both HERGs and LERGs live in clearly larger galaxies with larger black holes.
Our findings here broadly agree with what has been observed so far for the largest galaxies galaxies, but the overlap seen in the properties are significant enough to suggest that the dichotomy in the two populations are generally only restricted to the largest galaxies with the biggest black holes.

\subsection{HERGs and LERGs in galaxy scaling relations}
 
\begin{figure*}
\begin{center}
\includegraphics[width=\linewidth]{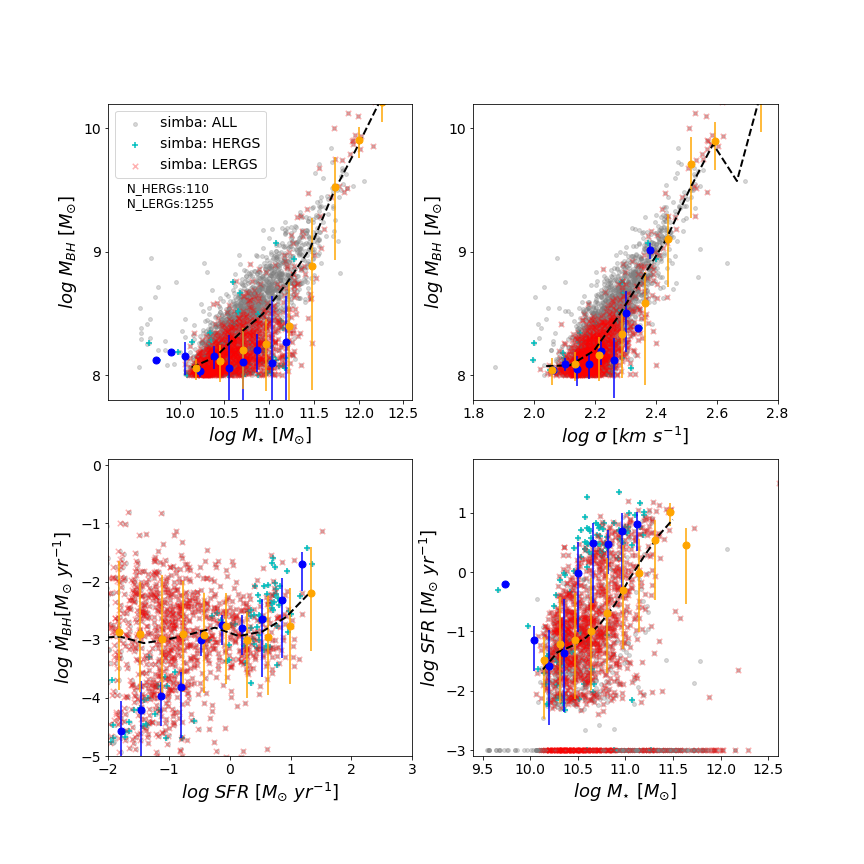}
\captionof{figure}{
Positions of HERGs (cyan plusses) and LERGs (red crosses) on black hole - galaxy scaling relations with respective blue and orange points showing the running median with error bars. From left to right: $\mbh-\mstar$, $\mbh-\sigma$, and BHAR-SFR relations. Grey points are all black hole hosts in \simba\ and the black dashed line is the median fit.}
\label{fig:SR}
\end{center}
\end{figure*}

For a complementary view of how RLAGN relate to the overall population, we examine the locations of HERGs and LERGs on key overall galaxy scaling relations. Particularly, we focus on the scaling relations between black hole mass and stellar mass or velocity dispersion, the black hole accretion rate vs. SFR, and sSFR vs. $\mstar$.

Figure~\ref{fig:SR} shows the position of HERGs and LERGs on these scaling relations. In each panel, blue pluses indicate the population of galaxies identified as HERGs while the red crosses show the population identified as LERGs. The small grey circles show all other resolved galaxies in \simba.
The blue and orange points with errorbars show the running medians for HERGs and LERGs respectively, while the black dashed line shows a median fit for all galaxies.

The upper left panel of Figure~\ref{fig:SR} shows the $\mbh$-$\mstar$ relation. This indicates that both HERGs and LERGs broadly lie on the same relation.  But compared to the overall population, interestingly both HERGs and LERGs lie at lower BH masses at a given $\mstar$ for \edit{$10^{10.5}\la (\mstar/\msun) \la 10^{11.5} $}, until one gets to the very highest LERG-dominated masses.  This indicates that RLAGN pick out galaxies that are growing their BHs comparatively quickly versus typical galaxies, catching up to the higher $\mbh$ typical for its $\mstar$.  Similarly albeit less dramatically, both HERGs and LERGs lie slightly below the $\mbh-\sigma$ relation, while LERGs dominate over HERGs at the high-$\mbh$, high-$\sigma$ region.  

For the BHAR-SFR relation (lower left panel), LERGs follow the median relation for all galaxies, while HERGs follow a steeper relation where they lie below the median trend at low-BHAR, low-SFR and cross over to slightly higher-BHAR, high-SFR at SFR$\sim1 \msun {\rm yr^{-1}}$.  This suggests that HERGs with the smallest SFRs are not growing their black holes particularly quickly.  Finally, when looking at the $\mstar$-SFR relation (lower right panel), we can also see a clear distinction between the two populations, with HERGs tend to lie above the median trend of the $\mstar$-SFR relation while LERGs lie along it.  Hence HERGs are preferentially picking out galaxies forming stars particularly quickly at their given stellar mass.

Overall, these results highlight that RLAGN are not randomly selected from the overall galaxy population.  RLAGN tend to have somewhat lower BH masses compared to the overall galaxy population at a given $\mstar$ or $\sigma$, except at the very highest and lowest $\mstar$.  Meanwhile, HERGs are differentiated from LERGs at low SFR, as well as showing higher sSFRs at a given $\mstar$.  These predictions can be tested with upcoming multi-wavelength radio surveys such as MIGHTEE.

\section{Conclusions}
\label{sec:conclusions}

We use the state-of-the-art \simba\ hydrodynamical galaxy formation simulation to examine the 1.4~GHz radio luminosity of galaxies, arising from both star formation and AGN activity. We define a population of RLAGN as those having full-powered jets in \simba, i.e. $0<\fedd<0.02$ in \simba's two-mode accretion model.  Within this a population we classify high- and low-excitation radio galaxies (HERGs and LERGs) based on their dominant mode of accretion (torque-limited vs. Bondi, respectively), based on the physically-motivated scenario that HERGs accrete predominantly from a cooler disk while LERGs are dominated by hot advection-dominated flow~\citep[e.g.][]{BestandHeckman2012}.  We examine basic demographics such as the 1.4~GHz radio luminosity function, and whether their radio emission is driven by star formation or black hole accretion.  We then study the global properties of RLAGN in relation to the overall galaxy population, as well as HERGs and LERGs, to understand whether there are distinct differences between their host galaxies. 

Considering the RLAGN population demographics, we find the following results:
\begin{itemize}
    \item \simba\ yields a total radio luminosity function (RLF) and an AGN-dominated RLF at $z=0$ in good agreement with observations by~\citet{MS2007}.   Star formation (SF) processes dominate the faint end of the RLF, while AGN dominates the bright end.  The cross over occurs at $\log \power \approx22.5 \WHz$, which is only slightly lower than that inferred from observations ($\log \power \approx23 \WHz$), likely within systematics. The agreement between RLAGN, as well as SF- and AGN-dominated sources in \simba\ and observations is a notable and novel success of the model, and suggests that \simba's accretion and jet feedback models are plausible.
    \item The RLF for LERGs is in quite good agreement with observations, except perhaps at the faintest luminosities where \simba\ predicts a rollover in the LERG RLF that is mostly beyond the range of current observations.  LERGs are dominantly found in quiescent galaxies. 
    \item The star forming and quiescent galaxies in \simba\ trace the SF and AGN RLFs observed by~\citet{MS2007} respectively, suggesting that star formation dominates the radio power in star forming galaxies while AGN dominates the radio power in quenched galaxies. In \simba, AGN jets play a critical role in the quenching of galaxies, so the broad agreement of \simba\ with data suggests this may also be the case in real galaxies.
\end{itemize}

Examining the RLAGN population in more detail via our differentiation between HERGs and LERGs, we conclude the following:
\begin{itemize}
    \item HERGs typically reside in hosts with mildly lower $\mstar$, lower $\mbh$, and higher sSFRs than that of LERGs but reside in halos of similar $\mhalo$ and $\sigma_{\star}$. Both populations of HERGs and LERGs lie toward the high end of the stellar- and black hole mass functions, while lying to the low end of the sSFR and $\fedd$ functions, relative to a mass-matched sample drawn from the galaxy overall population while spanning similar ranges in $\mhalo$ and $\sigma_{\star}$.
    \item HERGs are much more common in star-forming galaxies than LERGs, although both HERGs and LERGs appear among quenched galaxies.
    \item The most massive galaxies with the largest black holes in \simba\ are almost exclusively LERGs; 7 of the 9 most massive galaxies overall are LERGs.
    \item Despite these modest differences between the distributions of the two populations, the overlap in the HERG and LERG distributions is substantial.  \simba\ thus predicts that as one probes towards lower radio power with upcoming surveys, there becomes increasingly less differentiation between the host galaxy demographics of HERGs and LERGs. 
    \item Relative to the overall galaxy population, most HERGs and LERGs lie slightly below the median trend of the $\mbh-\mstar$ and $\mbh-\sigma$ relations, with a few HERGs at low $\mstar$ and $\sigma$.  LERGs lie along the $\mdot$--SFR relation, while HERGs lie below it at low SFRs crossing to above the median at SFR$\sim1 \msun {\rm yr^{-1}}$.  LERGs lie along the $\mstar$--sSFR main sequence, while HERGs typically have higher SFR at a given $\mstar$. 
\end{itemize}

These results demonstrate that \simba's black hole accretion and feedback schemes work together to produce a RLAGN population that is in reasonable agreement with observations.  As far as we are aware, this is the first time a cosmological galaxy formation simulation has reproduced such observations in this level of detail.  This opens up many potential areas of investigation associated with RLAGN, in order to understand their role in galaxy evolution.  With the emergence of next-generation radio surveys such as MIGHTEE, the evolution of the RLAGN population, both star formation-dominated and AGN-dominated, promises to play an important role in understanding how massive galaxies form, grow, and quench.

\section*{Acknowledgements}
The authors acknowledge helpful discussions with Philip Best and Imogen Whittam. We thank Philip Hopkins for making \gizmo\ public, and providing our group with early access. We thank Robert Thompson for developing {\sc Caesar}, and the {\sc yt} team for development and support of {\sc yt}.
NT acknowledges support from the South African Radio Astronomy Observatory, which is a facility of the National Research Foundation, an agency of the Department of Science and Technology. NT and RD acknowldedge support from Newton Mobility Grant NMG-R1-180195 from the U.K. Royal Society.
RD acknowledges support from Wolfson Research Merit Award WM160051 from the U.K. Royal Society.
DAA acknowledges support by NSF grant AST-2009687 and by the Flatiron Institute, which is supported by the Simons Foundation.
The \simba\ simulation was run on the DiRAC@Durham facility managed by the Institute for Computational Cosmology on behalf of the STFC DiRAC HPC Facility. The equipment was funded by BEIS capital funding via STFC capital grants ST/P002293/1, ST/R002371/1 and ST/S002502/1, Durham University and STFC operations grant ST/R000832/1. DiRAC is part of the National e-Infrastructure.

\section*{Data Availability}
The \simba\ data used to derive the findings in this study are publicly available at \url{http://simba.roe.ac.uk}, and analysis codes are available upon request.






\bibliographystyle{mnras}
\bibliography{RADIO_SIMBA.bib}

\appendix
\section{Varying the estimation of the 1.4 GHz radio luminosity}
\label{AA}
\begin{Figure}
\begin{center}
\includegraphics[width=\linewidth]{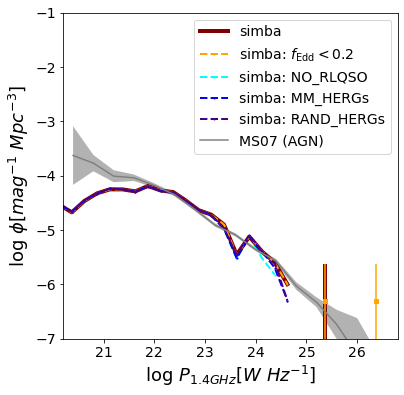}
\captionof{figure}{Variations of the AGN RLF. The maroon solid line shows the AGN RLF in \simba\ used throughout this work. The orange dashed line shows the AGN RLF if we were to select radio galaxies as those with $ \fedd < 0.2$. The cyan dashed line shows the AGN RLF without the additional radio power applied to radio galaxies to account for the RLQSO population. The blue dashed line shows the AGN RLF if we apply the additional radio luminosity to 10$\%$ of HERGs in the most massive dark matter halos as to account for the RLQSO population. The purple dashed line shows the AGN RLF if we apply the additional radio luminosity to a random 10$\%$ of HERGs to account for the RLQSO population. These are in comparison with the AGN contribution to the RLF from~\citet{MS2007}. }

\label{fig:RLF_var}
\end{center}
\end{Figure}
We consider variations to the AGN radio luminosity function based on our estimation of radio luminosities in \S\ref{sec:RLAGN}. In Figure~\ref{fig:RLF_var} we show the AGN RLFs for different methods for estimating the 1.4 GHz radio luminosity and compare these with the observed AGN RLF by \citet[grey band]{MS2007}. \newline
The maroon solid line shows the RLF as defined throughout this work and shown in Figure~\ref{fig:RLF}. \newline
\indent The orange dashed line shows the AGN RLF if we were to define galaxies as those with $\fedd<0.2$ which is when \simba\ switches to jet feedback mode, as opposed to $\fedd<0.02$ when jets reach full velocity. In this case, there is one additional galaxy increasing the radio galaxy count from from 1365 to 1366. We then estimate the radio luminosities in the same way as \S\ref{sec:radio_em} which results in one galaxy reaching a radio luminostiy of $\log\ P > 26 \WHz$ but which lies above the observed RLF. \newline
\indent The cyan dashed line shows the AGN RLF if we do not add any additional radio luminosity to make up for the RLQSO population not reproduced in \simba. This makes a minor difference to the RLF with the slight decrease at $\log P \gtrsim 24 \WHz $. There are also no sources reproduced at $\log\ P > 25 \WHz$. \newline 
\indent The blue dashed line shows the RLF if instead of applying the additional radio luminosity to 10$\%$ of HERGs that are centrals and in the most massive dark matter halos, we apply the additional luminosity to 10$\%$ of HERGs in the most massive dark matter halos, regardless of it being a central or satellite. We again see a negligible effect in that the only difference in a slight decrease in the RLF at $\log\ P \sim 24.5 \WHz $ and with no sources reproduced at $\log\ P \gtrsim 25 \WHz$.  \newline
\indent Finally, we apply the additional radio luminosity to a random 10$\%$ of HERGs shown by the purple dashed line. The only noticeable difference this makes is the lack of objects reproduced at $\log\ P \gtrsim 25 \WHz$. 
\newline 
\indent We can thus confirm that, other than reproducing the source at high luminosity, the manner in which we account for the RLQSO population makes negligible difference to our results. Similarly, defining radio galaxies as those with $\fedd<0.2$ rather than $\fedd<0.02$ makes little difference to our results and produces a source at high luminosity above the observed RLF. 
\section{Variation of the bondi fraction accretion timescale}
\begin{Figure}
\begin{center}
\includegraphics[width=\linewidth]{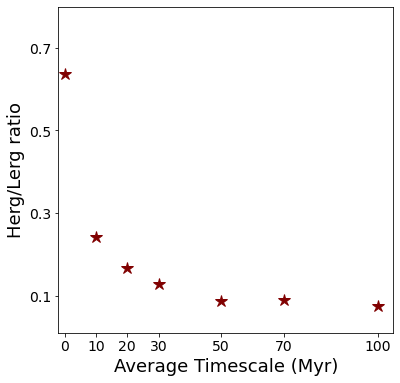}
\captionof{figure}{Ratio of HERGs to LERGs as a function of the timescale over which the average Bondi fraction, $f_{\rm Bondi}$, is computed.
}
\label{fig:time_ave}
\end{center}
\end{Figure}
Figure~\ref{fig:time_ave} shows the ratio of HERGs to LERGs due to the timescale over which the fraction of accretion in Bondi, $f_{\rm Bondi}$ has been averaged as per \S\ref{sec:herglerg}. That is, we take the average Bondi accretion rates of radio galaxies over 0 (instantaneous) to 100 Myr as to define the optimal timescale to average over the accretion stochasticity. 
\newline
\indent We note that there is an exponential decrease with longer timescales with the sharpest decrease from 0 to 10 Myr. The decrease tapers off between 30 and 50 Myr with no significant differences between 50, 70 and 100 Myr. \edit{In \S\ref{sec:herglerg} we thus} compute the average Bondi fraction over a 50 Myr timescale such that the ratio between HERGs and LERGs is roughly an order of magnitude. This is comparable to the dynamical time in the vicinity of the black hole, and averaging over longer timescales will make no significant difference to our results.

\bsp	
\label{lastpage}
\end{document}